\documentclass[12pt,a4paper]{article}
\usepackage{epsfig,amsmath,amssymb,subfigure,placeins}
\usepackage[colorlinks=true,linktocpage=true,linkcolor=blue,citecolor=blue]{hyperref}
\usepackage{color}
\newcommand{\jpsi}{J / \psi}

\newcommand{\old}[1]{}

\newcommand{\be}{\begin{equation}}
\newcommand{\ee}{\end{equation}}
\newcommand{\ba}{\begin{eqnarray}}
\newcommand{\ea}{\end{eqnarray}}
\newcommand{\bi}{\begin{itemize}}
\newcommand{\ei}{\end{itemize}}

\begin{document}
\begin{flushright}
{\normalsize
%June 21 2008}
}
\end{flushright}
\vskip 0.1in
\begin{center}
{\large{\bf Dissociation of Quarkonium in a Complex Potential }}
\end{center}
\vskip 0.1in
\begin{center}
Lata Thakur, Uttam Kakade and Binoy Krishna Patra\footnote{binoyfph@iitr.ac.in}\\
{\small {\it Department of Physics, Indian Institute of 
Technology Roorkee, India, 247 667} }
\end{center}
\vskip 0.01in
\addtolength{\baselineskip}{0.4\baselineskip} %wide line spacing

%opening

\begin{abstract}
We have studied the quasi-free dissociation of quarkonia through a
complex potential which is obtained by correcting both
the perturbative and nonperturbative terms of the $Q \bar Q$
potential at T=0 through the dielectric function in real-time formalism. 
The presence of confining nonperturbative term even above the transition 
temperature
makes the real-part of the potential more stronger and thus makes the quarkonia 
more bound and also enhances the (magnitude) imaginary-part 
which, in turn contributes more to the thermal width, compared to 
the medium-contribution of the perturbative term alone. These cumulative observations result the quarkonia to dissociate
at higher temperatures. Finally we extend our calculation
to a medium, exhibiting local momentum anisotropy, by calculating the
leading anisotropic corrections to the propagators in Keldysh
representation. The presence of
anisotropy makes the real-part of the potential stronger but the 
imaginary-part is weakened slightly. However, since the medium
corrections to the imaginary-part is a small perturbation to the 
vacuum part, overall the anisotropy makes the dissociation 
temperatures higher, compared to isotropic medium.
\end{abstract}
%\maketitle

PACS:~~ 12.39.-x,11.10.St,12.38.Mh,12.39.Pn

\vspace{1mm}
{\bf Keywords}: 
Quantum Chromodynamics, Debye mass, Momentum anisotropy, String tension, 
Dielectric permittivity, 
Quark-gluon plasma, Heavy quark potential, Decay width.

\section{Introduction}

The study of the heavy quarkonium states at finite temperature got 
impetus after the proposal by Matsui and Satz~\cite{Matsui} 
where the dissociation of quarkonium due to the color screening in 
the deconfined medium signals the formation of quark gluon
plasma (QGP)~\cite{bkpejc10}.
The assumption behind the proposal is that the 
medium effects can be envisaged through a temperature-dependent 
heavy quark potentials and have been studied over the decades either 
phenomenologically or through lattice based free-energy 
calculations~\cite{Mocsy08,Mocsy09}.
In recent years there have been important theoretical developments in 
heavy quarkonium physics where a sequence of effective field
theories (EFT)~\cite{Brambilla:1999xf, Brambilla:2004jw,Caswell,Thacker, Bodwin} are
derived by exploiting the hierarchy of different scales of heavy quark 
bound state: $m_{Q}\gg m_{Q}v\gg m_{Q}v^2 $, due to its large 
quark mass ($m_Q$). For example, the heavy quark system can be 
described by non-relativistic quantum chromodynamics (NRQCD) obtained 
from QCD by integrating out the mass. To describe the bound state of two 
quarks, one can further integrate out the typical momentum exchange ($m_Q v$)
between the bound quarks~\cite{Brambilla:1999xf, Brambilla:2004jw} and leads to potential 
non-relativistic QCD (pNRQCD) which describes a bound state by a two point 
function satisfying the Schr\"odinger equation through the 
potentials as the matching coefficients of the Lagrangian.
The EFT can also be generalized to finite temperature to justify the use of 
potential models at finite temperature \cite{Brambilla:2008cx} but
the thermal scales ($T$, $g T$ etc.) make the analysis complicated.
{\em For example}, when the binding energy is larger than the temperature, 
there is no medium modifications of the heavy quark 
potential~\cite{Brambilla:2008cx} but the properties of quarkonia states 
will be affected through the interactions with ultra-soft gluons. As a 
result the binding energy gets reduced and a finite thermal width 
is developed due to 
the medium induced singlet-octet transitions arising from the dipole 
interactions~\cite{Brambilla:2008cx}. This temperature regime is relevant 
for the $\Upsilon$(1S) suppression at the LHC.
In the opposite limit (binding energy $<T<gT$), the
potential acquires an imaginary component~\cite{Brambilla:2008cx}.
However beyond the leading-order, the above distinctions are no more
possible.

In non EFT, the heavy quark potential is defined from the 
late-time behavior of the Wilson loop \cite{Wilson:1974sk,Makeenko} and 
can be directly calculated either in Euclidean-time lattice simulations 
or in perturbation theory~\cite{Berges}. However the definition of the 
heavy-quark potential related to the finite-temperature real-time 
Wilson loop, as employed in the lattice QCD extraction is also 
based on an application of EFT~\cite{Barchielli:NPB296,Rothkopf:MPL28},
where the derivation proceeds on the level of NRQCD and happens to be
complex \cite{Laine:2006ns,Brambilla:2008cx}. The imaginary part of the 
potential can be interpreted as the Landau damping \cite{Beraudo:2007ky} which
describes the decaying of the $Q \bar Q$ correlation
with its initial state due to scatterings in the plasma.

The separation of thermal scales in EFT ($T\gg g T\gg g^2T$) (in weak-coupling 
regime), in practice is not evident and one needs lattice techniques to test the approach. 
To understand the color screening in the strong-coupling regime,
lattice calculations of the spatial correlation 
functions of static quarks are needed. 
In principle it is possible to study the problem of quarkonium 
dissolution without any use of potential models. 
Recently a lot of progress has been made in this direction in which 
the in-medium properties of different quarkonium states are encoded in 
spectral functions in terms of the Euclidean 
meson correlation functions constructed on the lattice~~\cite{Karsch,Mocsy05,Wong05,Mocsy06,Cabrera07,
Mocsy07,Alberico08,Mocsy08prd,Forcrand}.
However, the reconstruction of the spectral functions from the lattice meson 
correlators turns out to be
very difficult, and despite several attempts its outcome  
still remains inconclusive. One remarkable feature of the studies 
of the lattice meson
correlators is their feeble temperature dependence despite the expected 
color screening. This seems to be puzzling!

Not only is the determination of the effective potential still an open 
question but also there are other related issues
such as relativistic effects, thermal width of the states and
contribution from quantum corrections that need to be taken care of.
The physical picture of quarkonium dissociation in a medium has undergone a 
theoretical refinements over the last couple of years\cite{dpalejc00,bkpplb01,
bkpnpa02,bkpejc04,bkpejc05,bkpejc06}.  
Experimentally, the properties of thermally produced heavy quarkonium 
can be observed through the energy spectrum of their decay products
(dilepton pair)~\cite{Schenke,Martinez}. The dissociation of quarkonium 
resonances correspond to the disappearance of their peaks in the 
dilepton production rate. However, merely estimating the 
energy levels from the potential models does not allow one to
reconstruct the spectral function, which can determine the production 
rate \cite{Burnier07}. Physically a resonance dissolves 
into a medium through the broadening of its peak gradually, due to its
interaction with the partons in the medium.
Earlier it was thought that a quarkonium state is dissociated 
when the Debye screening becomes so strong that it inhibits the 
formation of bound states but nowadays a quarkonium is dissociated 
at a lower temperature \cite{Laine:2006ns,Burnier07} even though its binding
energy is nonvanishing, rather is overtaken by the Landau-damping 
induced thermal width \cite{Laine:2007qy}, obtained 
from the imaginary part of the potential. Its consequences 
on heavy quarkonium spectral functions \cite{Burnier07,Miao10}, 
perturbative thermal widths \cite{Laine:2007qy, Brambilla:2010vq}
quarkonia at finite velocity \cite{Escobedo:2011ie},
in a T-matrix approach 
\cite{Grandchamp:2005yw,Rapp:2008tf,Riek:2010py,Emerick:2011xu,Zhao:2011cv}, 
and in stochastic real-time dynamics \cite{Akamatsu:2011se} have 
been studied. Recently the dynamical evolution of the plasma 
was combined with the real and imaginary parts of the binding
energies to estimate the suppression of quarkonium \cite{Strickland:2011aa} in
RHIC and LHC energies.

As discussed above, in-medium corrections to the potential are
always accompanied with both real and imaginary components. In
the weak coupling regime, the Landau damping caused by the imaginary component
is the principal mechanism for the dissociation of heavy quark bound states.
Hence any realistic calculation of the spectral functions needs to 
incorporate both of the real and imaginary part. However the separation of the
scales, which are related to the screening of static electric fields ($gT$) 
and magnetic fields ($g^2T$) etc., are not satisfied at the strong-coupling 
limit and thus needs to handle nonperturbatively through the lattice 
studies. Although the lattice studies have shown that a sizable imaginary 
component is visible in the potential~\cite{Rothkopf12,Burnier12}
but may not be reliable because the necessary quality of the
data has not yet been achieved. One thus needs inadvertent support from the
potential models at finite
temperature as an important tool to complement the lattice studies.

Usually potential model studies are limited 
to the medium-modification of the perturbative part of the potential only.
It is found that the bulk properties of the QCD plasma phase, {\it e.g.}
the screening property, equation of state~\cite{Beinlich,Karsch00} 
etc. deviate from the perturbative  
predictions, even beyond the deconfinement temperature.
In the sequel, the phase transition in QCD for physical quark masses 
is found to be a crossover \cite{Aoki:2006we,phaseT}.  It is thus 
reasonable to assume that the string-tension does not vanish abruptly at 
the deconfinement point~\cite{string1,string2,string3}, so
one should study its effects on heavy quark potential even above $T_c$.
This issue, usually overlooked in the
literature where only a screened Coulomb potential was
assumed above $T_c$ and the linear/string term was neglected,
was certainly worth investigation.
Sometimes one-dimensional Fourier transform of the Cornell potential 
was employed with the assumption of color flux tube~\cite{dixit} in
one-dimension but at finite temperature, it may not be
the case since the flux tube structure may expand in more
dimensions~\cite{Satz}. Therefore, it would be better to consider the 
three-dimensional form of the medium modified Cornell potential.
% which has been done exactly in the present work.
Recently a heavy quark potential was obtained by correcting
both perturbative and nonperturbative terms in the 
Cornell potential, not its Coulomb part alone,
with a dielectric function encoding the effects of the deconfined
medium~\cite{prc_vineet}. The inclusion of nonvanishing string term,
apart from the Coulomb term made the potential more attractive
which can be seen by an additional long range Coulomb term, in addition
to the conventional Yukawa term.
In the short distance limit, the potential reduces to the vacuum one,
{\it i.e.}, the $Q \bar Q$ pair does not see the medium
% giving rise the duality between $V(r,T=0)$ and $V(0,T)$. 
whereas in the large distance limit, potential reduces to a 
long-range Coulomb potential with
a dynamically screened-color charge. Thereafter with this potential, the 
binding energies and
dissociation temperatures of the ground and the lowest-lying states of
charmonium and bottomonium spectra have been
determined~\cite{prc_vineet,chic_vineet}.

The discussions on the medium modifications of quarkonium 
properties referred above are restricted to isotropic medium only, it was until 
recently where the effect of anisotropy is considered in the heavy-ion
collisions~\cite{Romatschke:2003ms}. At the very early time of collision, asymptotic 
weak-coupling enhances the longitudinal expansion substantially than the 
radial expansion, thus the system becomes colder in the longitudinal direction
than in the transverse direction and causes an anisotropy in the 
momentum space. The anisotropy
thus generated affects the evolution of the system as well as the properties
of quarkonium states. In recent years, the effects of anisotropy on both
real and imaginary part of the heavy-quark potential
and subsequently on the dissociation of quarkonia states 
have been investigated in an anisotropic medium
\cite{Dumitru:2007hy,Dumitru:2009ni,Burnier:2009yu,Dumitru:2009fy,%
Margotta:2011ta} extensively.  
Recently we extended our aforesaid calculation~\cite{prc_vineet} for an
isotropic medium to a medium which
exhibits a local anisotropy in the momentum space by correcting the full Cornell
potential through the hard-loop resumed gluon propagator~\cite{lt}.
The presence of anisotropy introduces an angular dependence, in addition
to inter-particle separation, to the potential which
is manifested in weakening the screening of the potential. As a result
the resonances become more bound than in isotropic medium.
Since the weak anisotropy represents a perturbation to the
(isotropic) spherical potential, we obtained the first-order correction 
due to the small anisotropic contribution to the energy eigenvalues of
spherically-symmetric potential and explored how the properties of
quarkonium states change in the anisotropic medium. For example,
the dissociation temperatures are found minimum for the isotropic 
case and increase with the increase of anisotropy.

In the present work we aim to calculate the imaginary part, in addition
to the real part of the potential both in isotropic and anisotropic medium
by correcting the full Cornell potential, not its Coulomb part alone.
Therefore, we first revisit the leading
anisotropic corrections to the real and imaginary part of the retarded,
advanced and symmetric propagators through their self energies, and
then plug in their static limit to evaluate the real and imaginary part 
of the static potential, respectively.
This imaginary part provides a contribution to the width ($\Gamma$)
of quarkonium bound states~\cite{Laine:2006ns,Beraudo:2007ky,Laine:2007qy} which
in turn determines their dissociation temperatures by the
criterion: dissociation point of a particular resonance is defined as
the temperature where the twice of the (real part of) binding energy 
equals to $\Gamma$~\cite{Mocsy07,Burnier07,Escobedo,Laine} or from the
intersection of the real and imaginary part of the binding energies.
The structure of our paper is as follows. Section 2 is devoted
to the formalism of the potential 
in both isotropic and anisotropic medium. So we have started with 
a review of the retarded, advanced and symmetric propagators and 
self energies in Keldysh representation and their evaluation 
in HTL resummed theory in both isotropic and anisotropic medium 
in Section 2.1. With these ingredients, we calculate the real
and imaginary part of the (static) potential 
and subsequently studied the dissociation of charmonium and bottomonium
states
by calculating their real and imaginary binding energies and (thermal) widths 
for isotropic and anisotropic medium in subsection(s) 2.2 and 2.3,
respectively.
Moreover we have shown our results and try to explain them in terms of 
various effects: the contribution of the non-perturbative (string) term, 
the anisotropy, the screening scale etc.
Finally, we conclude our main results in Section 3.\\
%%%%%%%%%%%%%%%%%%%%%%%%%%%%%%%%%%%%%%%%%%%%
\section{Potential in a hot QCD medium}
As discussed earlier, any meaningful discussion of quarkonium properties
in thermal medium should include both real and imaginary parts for the
temperature-dependent potential. The hierarchy of scales assumed in
weak coupling EFT calculations may not be satisfied and the adequate 
quality of the data is not available in the present lattice calculations,
so one uses the potential model to circumvent the problem.

Because of the heavy quark mass ($m_Q$), the requirement: $m_Q \gg 
\Lambda_{QCD}$ and $T \ll m_Q$ is satisfied for the description of the 
interactions between a pair of heavy quarks 
and antiquarks at finite temperature, in terms of quantum mechanical 
potential. So we can obtain the medium-modification to the vacuum potential
by correcting its {\em both short and long-distance part} with a dielectric
function $ \epsilon(p)$ encoding the effect of deconfinement~\cite{prc_vineet}
\begin{eqnarray}
\label{defn}
V(r,T)&=&\int \frac{d^3\mathbf p}{{(2\pi)}^{3/2}}
 (e^{i\mathbf{p} \cdot \mathbf{r}}-1)~\frac{V(p)}{\epsilon(p)} ~,
\end{eqnarray}
where we have subtracted an $r$-independent term (to renormalize the heavy 
quark free energy) which is the perturbative free energy of quarkonium at 
infinite separation \cite{Dumitru09}.
The functions, $V(p)$ and $\epsilon(p)$ are the Fourier transform (FT) of 
the Cornell potential and the dielectric permittivity, respectively. 
To obtain the FT of the potential, we regulate both terms with the same 
screening scale. However in the framework of Debye-H\"{u}ckel theory, 
Digal et al.~\cite{Digal:EPJC43} employed different screening functions,
$f_c$ and $f_s$ for the Coulomb and string terms, respectively, to obtain
the free energy. \footnote{In another calculation, different 
scales for the Coulomb and linear pieces were also employed 
in~\cite{megiasind,megiasprd} to include non-perturbative
effects in the free energy beyond the deconfinement temperature through 
a dimension-two gluon condensate.}

At present, we regulate both terms by multiplying with an exponential 
damping factor and is switched off after the FT is evaluated.
This has been implemented by assuming $r$- as distribution
($r \rightarrow$ $r \exp(-\gamma r))$. The FT of
the linear part $\sigma r\exp{(-\gamma r)}$ is
\begin{eqnarray}
\label{eq-6-3}
-\frac{i}{p\sqrt{2\pi}}\left( \frac{2}{(\gamma-i p)^3}-\frac{2}{(\gamma+ip)^3}
\right).
\end{eqnarray}
After putting $\gamma=0$,  we obtain the FT of the linear term $\sigma r$
as,
\begin{equation}
\label{eq-6-4}
\tilde{(\sigma r)}=-\frac{4\sigma}{p^4\sqrt{2\pi}}.
\end{equation}
The FT of the Coulomb piece is straightforward, thus the FT of the full 
Cornell potential becomes
\begin{equation}
\label{vk}
{V}(p)=-\sqrt{(2/\pi)} \frac{\alpha}{p^2}-\frac{4\sigma}{\sqrt{2 \pi} p^4}.
\end{equation}
The dielectric permittivity will be calculated once the 
self energies and propagators are obtained in HTL resummation theory.

%%%%%%%%%%%%%%%%%%%%%%%%%%%%%%%%%%%%%%
\subsection{HTL self-energies and propagators}
The naive perturbative expansion, when applied to gauge fields, suffers from 
various singularities and even the damping rate
becomes gauge dependent~\cite{Thoma}. Diagrams which are of higher order in 
the coupling constant ($g$) contribute to leading order. These problems can be
partly avoided by using the hard thermal loop (HTL) resummation 
technique \cite{Braaten} to obtain the consistent results, which are 
complete to leading-order. At the same time the infrared 
behavior is improved by the presence of effective masses in the HTL 
propagators. The HTL technique has been shown to be equivalent to the 
transport approach~\cite{kelly94,Blaizot94} and is more
advantageous because it can be naturally extended to fermionic 
self-energies and to higher-order diagrams beyond the semiclassical 
approximation.

We shall now calculate the finite temperature self energies and propagator in 
real-time formalism~\cite{Carrington} where the propagators acquire a $2 \times 2 $ matrix
structure:
\begin{eqnarray}
 \label{2a4}
  D^0  =  \left (\begin{array}{cc} D_{11}^0 & D_{12}^0\\
                             D_{21}^0 & D_{22}^0\\
            \end{array} \right )~, 
\end{eqnarray}
where each component has zero and finite temperature part which contains
the distribution functions. In equilibrium, the distribution functions 
correspond to either (isotropic) Bose ($f_B$) or Fermi distribution 
($f_F$) function. Away from the equilibrium,
the distribution function needs to be replaced by the corresponding 
non-equilibrium one extracted from viscous hydrodynamics.
The nonequilibrium situation arises due to preferential expansion and non zero 
viscosity and as a consequence, a local anisotropy in momentum space sets in.
However, we consider a system close to equilibrium where the distribution
function can be obtained
from an isotropic one by removing particles with a large momentum-component
along the direction of anisotropy~\cite{Romatschke:2003ms,Martinez09}, ${\bf{n}}$, i.e., 
\begin{equation}
f_{\rm{aniso}} (\mathbf{p}) = f_{\rm{iso}}\left( \sqrt{\mathbf{p}^{2}
 + \xi(\mathbf{p}.\mathbf{n})^{2}}  \right)\\
 \approx f_{\rm{iso}}(p)\left[1-\xi \frac{({\bf {p.n}})^{2}}{2pT}
 (1\pm f_{\rm{iso}}(p))\right]~.
 \label{fun}
\end{equation}
The anisotropic parameter $\xi$ is related to the shear viscosity-to-entropy 
density ($\eta/s$) through the one-dimensional Navier Stokes formula by
\begin{equation}
\xi= \frac{10}{T \tau}~\frac{\eta}{s},~                 
\end{equation}
where $1/\tau$ denotes the expansion rate of the fluid element.
The degree of anisotropy is generically
defined by,
\begin{equation}
\xi = \frac{\langle \mathbf{k}_{T}^{2}\rangle}{2\langle k_{L}^{2}\rangle}-1~,~
\label{anparameter}
\end{equation}
where $ {k}_{L}= \mathbf{k}.\mathbf{n} $ and ${\bf k}_T = 
\mathbf{k}-\mathbf{n}(\mathbf{k}.\mathbf{n}) $ are the components of momentum
parallel and perpendicular to the direction of anisotropy, $\mathbf{n}$,
respectively. The positive and negative values of $ \xi$ correspond to the
squeezing and stretching of the distribution function in the direction of
anisotropy, respectively. However, in relativistic nucleus-nucleus
collisions, $\xi$ is found to be positive.
A useful representation of the propagators in real-time formalism is
the Keldysh representation where the linear combinations of four
components of the matrix, of which only three are independent, give the
relation for the retarded (R), advanced (A) and symmetric (F) propagators,
respectively :
\begin{eqnarray}
\label{2a6}
   D_R^0 = D_{11}^0 - D_{12}^0 ~,~ D_A^0 = D_{11}^0 - D_{21}^0 ~,~
   D_F^0 = D_{11}^0 + D_{22}^0  ~.
\end{eqnarray}
Only the symmetric component involves the distribution functions 
and is of particular advantage for the
HTL diagrams where the terms containing distribution
functions dominate. The similar relations for 
the self energies are :
\begin{eqnarray}
\Pi_R  =  \Pi_{11}+\Pi_{12} ~,~
\Pi_A  =  \Pi_{11}+\Pi_{21} ~,~
\Pi_F  =  \Pi_{11}+\Pi_{22} ~. \label{2a13}
\end{eqnarray}
Resumming the propagators through the Dyson-Schwinger equation, 
the retarded (advanced) and symmetric propagators can be written as 
\begin{eqnarray}
 {D}_{R,A}&=&D_{R,A}^0+D_{R,A}^0\Pi_{R,A}{D}_{R,A}~, \label{2b2}\\
 {D}_{F}&=&D_{F}^0+D_{R}^0\Pi _R{D}_{F}+D_F^0\Pi_{A} {D}_{A}+ 
 D_{R}^0\Pi _{F}{D}_{A}~. \label{2b7}
\end{eqnarray}
Substituting the symmetric propagator $D_F^0(P)$ in terms of the retarded 
and advanced propagator, the resummed symmetric propagator can be 
expressed as 
\begin{eqnarray}
D_{F}(P)= && (1+2f_B)\, \mbox{sgn}(p_0)\,
\left[D_{R}(P)-{D}_{A}(P)\right]
\nonumber \\
&& +D_{R}(P)\,\left[\Pi _F(P)-(1+2f_B)\, \mbox{sgn}(p_0)\, [\Pi
_R(P)-\Pi _A(P)]\right] \,  D_A(P)~. \label{2b8}
\end{eqnarray}
To calculate the static potential in isotropic medium, only the temporal 
component (L) of the propagator is needed so the retarded (advanced) 
propagator in the simplest Coulomb gauge can be written as 
\begin{eqnarray}
 D^L_{R,A(iso)}=D^{L(0)}_{R,A}+D^{L(0)}_{R,A}\Pi^L_{R,A(iso)}{D}^L_{R,A(iso)}~. \label{3b2}
\end{eqnarray}
So far the resummation is done in isotropic medium, however we now
extend it in a medium which exhibits a weak anisotropy ($\xi
\ll 1$). Therefore we first expand the propagators and self 
energies around isotropic limit and retain only the linear term:
\begin{equation}
 D=D_{\rm{iso}}+\xi D_{\rm{aniso}},\,\,\,\,\,\,\,\Pi=\Pi_{\rm{iso}}+\xi
\Pi_{\rm{aniso}} ~.\label{3b4}
\end{equation}
Thus in the presence of small anisotropy, the temporal component of the 
retarded (advanced) propagator becomes 
\begin{equation}
D^L_{R,A(aniso)}= D^{L(0)}_{R,A}\, \Pi_{R,A (aniso)}^L{D}^{ L}_{R,A (iso)}+D_{R,A}^{L(0)}\,
 \Pi_{R,A (iso)}^L{D}^ L_{R,A (aniso)}\label{3b6}
\end{equation}
whereas with the notations for the difference of propagators and 
self-energies : $\Delta D^L_{RA(aniso)}=[{D }^L_{R(aniso)}(P)-{D}^L_{A(aniso)}(P)]$,
$\Delta D^L_{RA(iso)}=[{D }^L_{R(iso)}(P)-{D}^L_{A(iso)}(P)]$,
$\Delta \Pi^L_{RA(aniso)}=[\Pi^L _{R(aniso)}(P)-\Pi^L _{A(aniso)}(P)]$, and
$\Delta \Pi^L_{RA(iso)}=[\Pi^L _{R(iso)}(P)-\Pi^L _{A(iso)}(P)]$,
the symmetric propagator can be obtained~\cite{Dumitru09}, 
\begin{eqnarray}
{D}^L_{F(aniso)}(P)&= & (1+2f_{B(iso)})\, \mbox{sgn}(p_0)\, 
\Delta^L_{RA(aniso)}+2f_{B(aniso)}\,
\mbox{sgn}(p_0)\, \Delta^L_{RA(iso)} +{D}^L_{R(iso)}(P)\,\left[\Pi _{F(aniso)}^L(P)
\right. \nonumber\\
&&\left. -(1+2f_{B(iso)})\,
\mbox{sgn}(p_0)\, \Delta \Pi^L_{RA(aniso)} \ -2f_{B(aniso)}\,
\mbox{sgn}(p_0)\, \Delta \Pi^L_{RA(iso)} \right] \, {D}^L_{A(iso)}(P)~.
 \label{3b7}
\end{eqnarray}
To solve the propagators, we will now calculate the gluon self energy from
the quark and gluon loops. The contribution 
of the quark loop \cite{Dumitru09} to the self energy with external 
and internal momenta as $P(p_0,{\mathbf p})$ and $K(k_0,{\mathbf k})$,
respectively (with $Q=K-P$):
\begin{eqnarray}
\Pi^{\mu\nu}(P)=-\frac{i}{2}N_{f}g^{2}\int\frac{d^4 K}{(2\pi)^4}
 tr[\gamma^{\mu}S(Q)\gamma^{\nu}S(K)]
\label{en} 
\end{eqnarray}
gives the retarded self energy  
\begin{eqnarray}
\Pi_{R}^{\mu\nu}(P) =-\frac{i}{2}N_{f}g^{2}\int\frac{d^4 K}{(2\pi)^4}
 \left( tr[\gamma^{\mu}S_{11}(Q)\gamma^{\nu}S_{11}(K)]-
 tr[\gamma^{\mu}S_{21}(Q)\gamma^{\nu}S_{12}(K)]\right).
\label{en1} 
\end{eqnarray}
Redefining the fermionic propagators:
$S_{\rm{R,A,F}}(K)\equiv \displaystyle{\not} K \tilde \Delta_{R,A,F}(K)$,
the longitudinal-part of the self energy becomes, in the limit of massless
quarks
\begin{eqnarray}
\Pi_{R}^{L}(P)=-iN_{f}g^{2}\int\frac{d^4 K}{(2\pi)^4}(q_0k_0+{\bf q.k})
 \left[\tilde{\Delta}_F(Q)\tilde{\Delta}_R(K)+\tilde{\Delta}_A(Q)
 \tilde{\Delta}_F(K)\right. \nonumber\\
 \left.+\tilde{\Delta}_A(Q)\tilde{\Delta}_A(K)+\tilde{\Delta}_R(Q)
\tilde{\Delta}_R(K)\right].
\label{en2} 
\end{eqnarray}
In the weak-coupling limit, the internal momentum ($T$) is 
much larger than the external momentum ($gT$), so the retarded self energy 
in the HTL-approximation simplifies into~\cite{Dumitru09}
\begin{eqnarray}
\Pi_{R}^{L}(P)=\frac{4\pi N_{f}g^{2}}{(2\pi)^4}\int kdk\int d\Omega 
f_{F}({\bf k})\frac{1-{(\hat{\bf k} \cdot \hat{\bf p})}^2}{\bf{(\hat k.\hat p}+
\frac{p_{0}+i\epsilon}{p})^2}~.
\end{eqnarray} 
After convoluting the distribution function, $f_F$ for quarks in an
(weakly) anisotropic medium from (\ref{fun}) 
%\begin{eqnarray}
%f_{F}({\bf k})=n_{F}(k)-\xi n_{F}^{2}(k)\frac{e^{k/T}}{2kT}({\bf {k.n}})^{2},
%\label{ex}
%\end{eqnarray}
the retarded quark self energy becomes 
\begin{eqnarray}
\Pi_R^L(P)=\frac{g^2}{2 \pi^2} N_f \sum\limits_{i=0,1} \int_{0}^{\infty} k\,
\Phi_{(i)}(k) d k \int_{-1}^{1} \Psi_{(i)}(s) d s ~, \label{3b11}
\end{eqnarray}
with
\begin{eqnarray}
\label{3b12}
\Phi_{(0)}(k)&=&n_F(k)\, ,\nonumber \\
\Phi_{(1)}(k)&=& - \xi n_F^2(k)\, \frac{k~e^{k/T}}{2 T}\, ,\nonumber \\
\Psi_{(0)}(s)&=&\frac{1-s^2}{(s+\frac{p_0+i\epsilon}{p})^2}\, ,\nonumber \\
\Psi_{(1)}(s)&=& \cos^2\theta_p\,
\frac{s^2(1-s^2)}{(s+\frac{p_0+i\epsilon}{p})^2}+\frac{\sin^2\theta_p}{2}
\frac{(1-s^2)^2}{(s+\frac{p_0+i\epsilon}{p})^2}~.
\end{eqnarray}
Here, the angle ($\theta_p$) is between ${\bf{n}}$ and ${\bf{p}}$ and
$s\equiv {\hat{\bf k}}\cdot {\hat{\bf p}}$. After decomposing 
into isotropic ($\xi$=0) and anisotropic ($\xi \ne 0$)
pieces, the isotropic and anisotropic terms become 
\begin{eqnarray}
 \Pi_{R(iso)}^L(P)&= &N_f \frac{g^2 T^2}{6}\left (\frac{p_0}{2 p}\ln 
\frac{p_0+p+i\epsilon} {p_0-p+i\epsilon} -1\right)~\label{3b13}\\
\Pi_{R(aniso)}^L(P)&=& N_f \frac{g^2 T^2}{6}\left(\frac{1}{6}+\frac{ 
\cos 2\theta_p}{2}\right)+
\Pi_{R(iso)}^L(P)\left(\cos 2\theta_p -\frac{p_0^2}{2 p^2} \left(1+3
\cos 2\theta_p \right)\right)\label{3b14},
\end{eqnarray}
respectively. In HTL-limit, the structure of gluon-loop contribution is 
the same as the quark-loop, apart from the degeneracy factor and
distribution function, so the quark and gluon loops together give the 
isotropic part of retarded (advanced) self-energy
\begin{eqnarray}
\Pi^{L}_{R,A(iso)}(P)=m_D^2\left(\frac{p_{0}}{2p}\ln\frac{p_{0}+p\pm i\epsilon}{p_{0}-p\pm i\epsilon}-1\right)~,
\label{iso}
\end{eqnarray}
with the prescriptions $+i\epsilon $ ($ -i\epsilon $), for the retarded 
(advanced) self-energies, respectively whereas the anisotropic part for
the retarded (advanced) self energies are
\begin{eqnarray}
\Pi^{L}_{R,A(aniso)}(P)=\frac{m_D^2}{6}\left(1+\frac{3}{2}\cos 2\theta_p \right)
+\Pi_{R(iso)}^{L}(P)\left(\cos(2\theta_p)-\frac{{p_{0}}^{2}}{2p^{2}}
(1+3\cos 2\theta_p)\right),
\label{aniso}
\end{eqnarray}
where $m_D^2$ (= $\frac{g^2 T^2}{6}(N_f+2 N_c$)) is the square of Debye mass. 

Similarly the isotropic and anisotropic terms for the temporal component
of the symmetric part are given by
\begin{eqnarray}
&&\Pi^{L}_{F(iso)}(P)=-2\pi i m_D^2\frac{T}{p}\Theta(p^2-{p_0}^2)~,\nonumber\\
&&\Pi^{L}_{F(aniso)}(P)=\frac{3}{2}\pi i m_D^2\frac{T}{p}
\left(\sin^2\theta_p 
+\frac{p_0^2}{{p}^2} (3\cos^2\theta_p-1)\right)~\Theta(p^2-{p_0}^2).
\label{sym}
\end{eqnarray}
Thus the gluon self-energy is found to have both real and imaginary 
part which are responsible for the Debye screening and the Landau damping,
respectively where the former is usually obtained from the retarded and 
advanced self energy and the later is obtained from the symmetric self energy alone. 

So, to evaluate the real part of the static potential, the real part of 
the temporal component of retarded (or advanced) propagator
(in static limit) is needed
\begin{eqnarray}
Re D^{00}_{R,A}(0,p)=-\frac{1}{(p^2+m_D^2)}
+\xi \frac{m_D^2}{6(p^2+m_D^2)^2}\left(3\cos 2\theta_p-1 \right),
\label{rtrdprop}
\end{eqnarray}
while for the imaginary part of the potential, the imaginary part of 
the temporal component of symmetric propagator is given by
\begin{eqnarray}
Im D^{00}_F (0,p)=\frac{-2\pi T m_D^2}{p(p^2+m_D^2)^2}
+\xi\left(\frac{3\pi T m_D^2}{2p(p^2+m_D^2)^2}\sin^2{\theta_p}
-\frac{4\pi T m_D^4}{p(p^2+m_D^2)^3} \left(\sin^2\theta_p-\frac{1}{3}\right)\right).
\label{f00}
\end{eqnarray}
With these real and imaginary part of the self energies and propagators,
we will obtain the (complex) potential in subsection(s) 2.2 and 2.3 for 
isotropic and anisotropic medium, respectively.
%%%%%%%%%%%%%%%%%%%%%%%%%%%%%%%%%%%%%%%%%%%%%%%%%%%%%%%%%%%
\subsection{Potential in isotropic medium}
\subsubsection{Real Part of the Potential}
The real part of the static potential can thus be obtained 
from eq.(\ref{defn}) by substituting the dielectric permittivity
 $\epsilon(p)$ in terms of the physical ``11"- component of the gluon propagator.
The relation between the dielectric permittivity and the static limit of 
the ``00"-component of gluon propagator in Coulomb gauge is obtained 
from the linear response theory:
\begin{equation}
\epsilon^{{}^{-1}}(p)=-\lim_{\omega \to 0} {p^2} D_{11}^{00}(\omega, p)~,
\label{ephs}
\end{equation}
where the propagator $D_{11}^{00}$ can be separated into real and imaginary 
parts as
\begin{eqnarray}
D^{00}_{11}(\omega,p)= Re D^{00}_{11}(\omega,p)+Im D^{00}_{11}(\omega,p).
%\frac{1}{2} \left(D^{00}_{R}+D^{00}_{A}+D^{00}_{F}\right),
\label{F1}
\end{eqnarray}
The real and imaginary parts can be further recast in terms retarded/advanced 
and symmetric parts, respectively
\begin{eqnarray}
Re D^{00}_{11}(\omega,p)= \frac{1}{2}\left( D^{00}_{R}+D^{00}_{A}\right)
\label{R}~~ {\rm{and}}~~
Im D^{00}_{11}(\omega,p)= \frac{1}{2} D^{00}_{F}.
\label{F}
\end{eqnarray}
Thus using the real part of retarded (advanced) propagator in isotropic medium 
\begin{equation}
Re D^{00}_{R,A}(0,p)=-\frac{1}{(p^2+m_D^2)}~,
\end{equation}
the real-part of the dielectric permittivity (also given in \cite{Schneider:prd66,
Weldon:1982,Kapusta})
becomes
\begin{equation}
\epsilon (p)=\left( 1 + \frac{m_D^2}{p^2} \right)~.
\label{dielectriciso}
\end{equation}
Note that this one-loop result in the linear response theory is a 
perturbative one, where the linear approximation in QCD holds as long 
as the mean-field four potential ($A_\mu^a$) is much smaller that the 
temperature~\cite{Blaizot:PR359}. Actually for the soft scales, the mean-field four potential
is at the order of $\sqrt{gT}$ and the linear approximation holds in the
weak-coupling limit.

However, if one assumes nonperturbative effects such as the string tension 
survive even much above the deconfinement point then the dependence of 
the dielectric function on the Debye mass may get modified.
So there is a {\em caveat} about the
validity of the linear dependence of the dielectric function ($\epsilon$)
on the square of the Debye mass $m^2_D$. For the sake of simplicity,
we put the remnants of the nonperturbative effects beyond the deconfinement
temperature by a multiplication factor $1.4$ to the leading-order Debye mass,
to take into account the next-to-leading corrections~\cite{Kajantie97} (the factor
is also obtained by fitting 
with the lattice results for the color-singlet free energy~\cite{Petreczky:2005bd}).
\begin{eqnarray}
\label{repot}
Re V_{(iso)}(r,T)&=&\int \frac{d^3\mathbf p}{{(2\pi)}^{3/2}}
 (e^{i\mathbf{p} \cdot \mathbf{r}}-1)~\left(-\sqrt{(2/\pi)} \frac{\alpha}{p^2}-\frac{4\sigma}{\sqrt{2 \pi} p^4}\right)\left( \frac{p^2}{(p^2+m_D^2)}\right) \nonumber\\
&\equiv & Re V_{1(iso)}(r,T)+Re V_{2(iso)}(r,T),
\end{eqnarray}
where $ Re V_{1(iso)}(r,T) $ and $Re V_{2(iso)}(r,T)$ correspond to the
medium modifications to the Coulomb and string term, respectively. After
performing the momentum integration, the Coulomb term becomes
\begin{equation}
Re V_{1(iso)}(r,T) =-\alpha m_D\left(\frac{e^{-\hat r}}{\hat r}+1\right)
\end{equation}
and the string term simplifies into
\begin{equation}
Re V_{2(iso)}(r,T) =\frac{2\sigma}{m_D}\left( \frac{(e^{-\hat r}-1)}{\hat r}+1\right).
\end{equation}
The real part of the potential in isotropic medium becomes (with $\hat{r}=rm_D$)
\begin{eqnarray}
\label{pis}
Re V_{(iso)}(\hat{r},T)=\left(\frac{2\sigma}{m_D}-\alpha m_D\right)\frac{e^{-\hat r}}{\hat r}
-\frac{2\sigma}{m_D \hat r}+\frac{2\sigma}{m_D}-\alpha m_D~,
\end{eqnarray}
which is found to have an additional long range Coulomb term, in addition 
to the conventional Yukawa term.
%%%%%%%%%%%%%%%%%%%%%%%%%%%%%%%%%%%%%%%%%%%
%%%%%%%%%%%%%%%%%%%%%%%%%%%%%%%%%%%%%%%%%%%%%%%%%fig1%%%%%%%%%%
\begin{figure}
\vspace{0mm}
\begin{subfigure}{
\includegraphics[width=5.9cm,height=6.5cm]{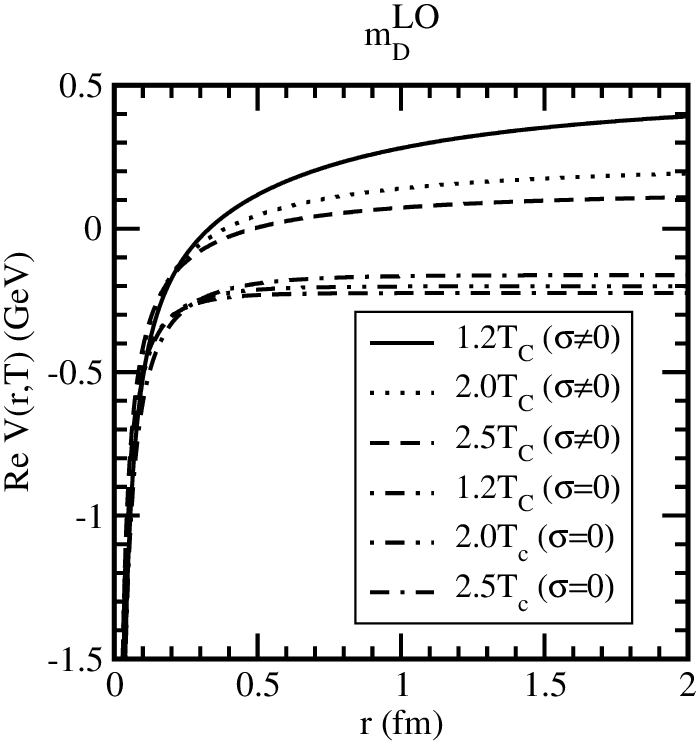} % Here is how to import EPS art
%\label{jpsi}
}
\end{subfigure}
\hspace{5mm}
\begin{subfigure}{
\includegraphics[width=5.9cm,height=6.5cm]{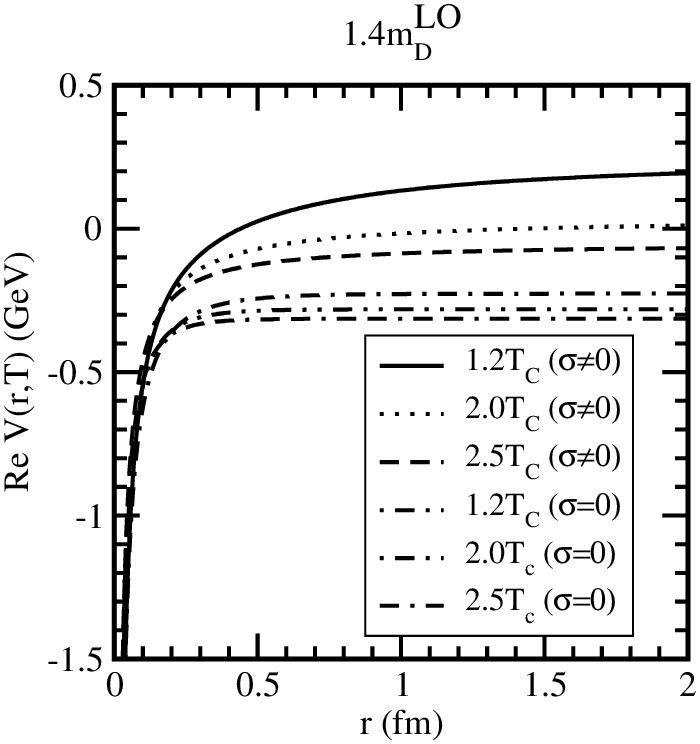}
%\label{upsi}
}
\end{subfigure}
\caption{\footnotesize The real-part of the static potential 
with ($ \sigma=0 $) and without ($\sigma \neq 0$) non-perturbative 
term in the potential. The left (right) panel of the figure denote 
the results obtained with the leading-order and lattice-fitted 
Debye masses, respectively.}
\vspace{0mm}
\label{jj}
\end{figure}
%%%%%%%%%%%%%%%%%%Figure 2 for potential%%%%%%%%%%%%%%%%%
In the small-distance limit ($\hat{r}\ll 1$), the above potential reduces to
the Cornell potential, {\it i.e.} the $Q \bar Q$-pair does not see the medium.
On the other hand, in the long-distance limit ($\hat{r} \gg1$), the potential
is simplified into, with high temperature approximation (ie. $\sigma/m_D(T)$
can be neglected): 
\begin{eqnarray}
\label{lrp}
{Re V_{(iso)}(r,T)} \approx -\frac{2\sigma}{m^2_{{}_D}r}
-\alpha m_{{}_D},
\end{eqnarray}
which, apart from a constant term, is Coulomb-like potential
by identifying $2 \sigma/m_{{}_D}^2$ with the square of the strong coupling
($g^2$). However if we compare the asymptotic limit ($r \rightarrow \infty$)
of our result (39) with the Digal et. al
\begin{eqnarray*}
F^{\rm{Digal}} (\infty,T)&=&
\frac{\Gamma(1/4)}{2^{3/2}\Gamma(3/4)}\frac{\sigma}{m_{{D}}(T)}-\alpha m_{{D}}(T)\\
F^{\rm{Our}}(\infty,T)&=& \frac{2\sigma}{m_{{D}}(T)}-\alpha m_{{D}}(T),
\end{eqnarray*}
the difference will be seen only in the string term only and may be due to 
the treatment of the problem classically or quantum mechanically. If we 
compare them quantitatively (with the Debye mass $m_D$= $ 1.4~m_D^{\rm{LO}}$),
the difference becomes tiny.

To see the effect of the linear term on the potential, in addition to the 
Coulomb term, we have plotted the (real-part) potential (in Fig.\ref{jj}) with 
($\sigma \neq 0$) and without string term ($\sigma=0$). We found that the 
inclusion of the linear term makes the potential attractive, compared to
potential with the Coulomb term only. Furthermore, to see the effects of the 
screening scale, we have also computed the potential with the Debye mass 
in next-to-leading order (1.4$ m_D^{\rm{LO}}$) which is 
seen less stronger than the leading-order result.
To see the effects of medium on the potential at T=0, we have evaluated 
the potential at different temperatures 
{\em viz.} at $1.2T_c$, $2.0T_c $ and $2.5T_c$, where the potential 
is found to decrease with the temperature at large distances 
and becomes short-range. Thus the deconfinement 
is reflected clearly in the
large-distance behavior of heavy quark potential at finite temperature, 
where the screening is operative. Thus the in-medium behavior of heavy 
quark bound states is used to probe the state of matter in QCD thermodynamics.
{\footnote{The real part of the singlet potential indeed coincides with the 
leading-order result of the so-called singlet free 
energy \cite{Petreczky:2005bd} because it contain entropy contribution.}}
%%%%%%%%%%%%%%%%%%%%%%%%%%%%%%%%%%%%%%%%%%%%%%%%%%%%%%%%%%%%%%
\subsubsection{Imaginary Part of the Potential: Thermal Width, 
$\Gamma_{\rm{iso}}$ }
The imaginary part of the potential originates from the static limit of 
symmetric self energy. Cutting rules at finite temperature allows one to
obtain the imaginary part by cutting open one of the hard thermal loop 
of the HTL propagator which represents physically the inelastic scattering
of the off-shell gluon off a thermal 
gluon~\cite{Brambilla:2008cx,Beraudo:2007ky,Laine:2007qy,Escobedo},
i.e. $g+ (Q \bar Q) \rightarrow g + Q +{\bar Q}$.
%between the light fermions of the hot medium and the heavy quarks. 
The imaginary part of the potential plays an important role in weakening 
the bound state peak or transforming it to mere threshold enhancement.
It leads to a finite width ($\Gamma $) for the resonance 
peak in the spectral function, which, in turn, determines the 
dissociation temperature. 
Dissociation is expected to occur while the (twice) binding energy decreases with 
the temperature and becomes equal to $\sim \Gamma$~\cite{ Mocsy07,Burnier07}.

To obtain the imaginary part of the potential in isotropic medium, we write
the temporal component of the symmetric propagator from (\ref{f00}) for
$\xi=0$, in the static limit, 
\begin{equation}
Im D^{00}_{F(iso)}(0,p)=\frac{-2\pi T m_D^2}{p(p^2+m_D^2)^2}.
\label{isof}
\end{equation}
However the same (\ref{isof}) could also be obtained for 
partons with space-like momenta ($p_{0}^2 < p^2$) from the retarded 
(advanced) self energy (\ref{3b13}), using the 
relation~\cite{Beraudo:2007ky,Burnier:2009yu}: 
\begin{eqnarray}
\ln\frac{p_{0}+p\pm i\epsilon}{p_{0}-p\pm i\epsilon}=
\ln | \frac{p_{0}+p}{p_{0}-p}| \mp i \pi 
\theta(p^2-{p_0}^2) ~.
\end{eqnarray}
Thus the imaginary part of the symmetric propagator (\ref{isof}) gives
the imaginary part of the dielectric function in isotropic medium :
\begin{eqnarray}
\epsilon^{-1} (p)= -\pi T m_D^2 \frac{p^2}{p(p^2+m_D^2)^2}.
\end{eqnarray}
One can then similarly find the imaginary part of the potential from the 
definition of potential (\ref{defn}) 
\begin{eqnarray}
Im V_{(iso)}(r,T)&=&-\int \frac{d^3\mathbf{p}}{(2\pi)^{3/2}}
(e^{i\mathbf{p} \cdot \mathbf{r}}-1)
\left(-\sqrt{\frac{2}{\pi}}\frac{\alpha}{p^2}-\frac{4\sigma}{\sqrt{2\pi}p^4}\right)
{p^2}\left[\frac{-\pi T m_D^2}{p(p^2+m_D^2)^2}\right]\nonumber\\
&\equiv& Im V_{1(iso)}(r,T)+Im V_{2(iso)}(r,T)~,
\end{eqnarray}
where $Im V_{1(iso)} (r,T)$ and $Im V_{2(iso)}
(r,T)$ are the imaginary parts of the potential due to the 
medium modification to the short-distance and 
long-distance terms, respectively:
\begin{eqnarray}
Im V_{1(iso)}(r,T)&=&-\frac{\alpha}{2\pi^{2}}\int d^{3}\mathbf{p} 
(e^{i\mathbf{p} \cdot \mathbf{r}}-1)
\left[\frac{\pi T m_D^2}{p(p^2+m_D^2)^2}\right],\nonumber\\
Im V_{2(iso)}(r,T)&=&-\frac{4\sigma}{({2\pi})^2}\int \frac{d^3 {\bf p}}
{(2\pi)^{3/2}}(e^{i {\bf p} \cdot {\bf r} }-1)
\frac{1}{p^2}
\left[\frac{\pi T m_D^2}{p(p^2+m_D^2)^2}
\right].
\end{eqnarray}
After performing the integration, the contribution due to the short-distance
term to imaginary part becomes (with $z=p/m_D$)
\begin{eqnarray}
Im V_{1(iso)}({\bf r},T)&=&-2\alpha T\int_0^{\infty} \frac{dz}{(z^2+1)^2}
\left(1-\frac {\sin{z\hat r}}{z\hat r}\right)\nonumber\\
&\equiv &-\alpha T\phi_0(\hat{r}),
\label{imiso1}
\end{eqnarray}
and the contribution due to the string term becomes
\begin{eqnarray}
Im V_{2(iso)}(r,T) &=&\frac{4\sigma T}{m_D^2}\int_0^{\infty} 
\frac{dz}{z(z^2+1)^2}
\left(1-\frac {\sin{z\hat r}}{z\hat r}\right)\nonumber\\
&\equiv &\frac{2\sigma T}{m_D^2}\psi_0(\hat{r})~,
\label{imiso1}
\end{eqnarray}
where the functions, $\phi_0(\hat{r})$ and $\psi_0(\hat{r})$ 
at leading-order in $\hat{r}$ are
\begin{eqnarray}
&&\phi_0(\hat{r})=-\alpha T\left(-\frac{{\hat{r}}^2}{9}
\left(-4+3\gamma_{E}+3\log\hat{r}\right)\right) \\
&&\psi_0(\hat{r})=\frac{\hat r^2}{6}+\left(\frac{-107+60\gamma_E 
+60\log(\hat r)}{3600}\right)\hat r^4+O(\hat r^5).
\end{eqnarray}
In the short-distance limit ($ \hat{r} \ll 1$), both the 
contributions, at the leading logarithmic order, reduce to
\begin{eqnarray}
&&Im V_{1(iso)}(r,T)=-\alpha T\frac{{\hat r^2}}{3}\log(\frac{1}{\hat r}),\\
\label{im1}
&&Im V_{2(iso)}(r,T)=-\frac{2\sigma T}{m_D^2}\frac{{\hat r^4}}{60}
\log(\frac{1}{\hat r}),
\label{im2}
\end{eqnarray}
thus the sum of Coulomb and string term gives the imaginary part 
of the potential in isotropic medium:
\begin{eqnarray}
Im V_{\rm{(iso)}} (r,\xi,T)=-T\left(\frac{\alpha {\hat r^2}}{3}
+\frac{\sigma {\hat r}^4}{30m_D^2}\right)\log(\frac{1}{\hat r}).
\label{imis}
\end{eqnarray}
One thus immediately observes that for small distances the imaginary part 
vanishes and its magnitude is larger than the case where only the Coulombic 
term is considered~\cite{Dumitru09} and thus enhances the width of the 
resonances in thermal medium.

\begin{figure}
\vspace{0mm}
\begin{subfigure}{
\includegraphics[width=5.9cm,height=6.5cm]{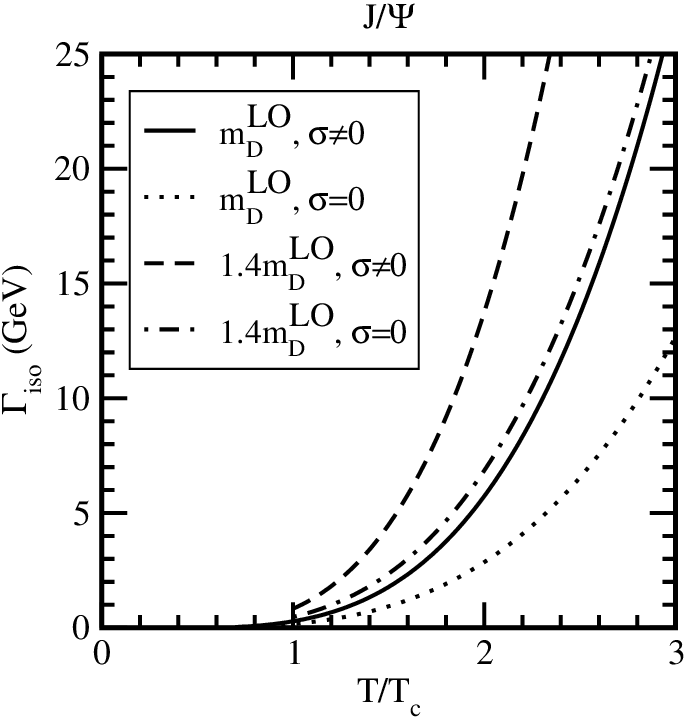} % Here is how to import EPS art
\label{gamaiso}}
\end{subfigure}
\hspace{5mm}
\begin{subfigure}{
\includegraphics[width=5.9cm,height=6.5cm]{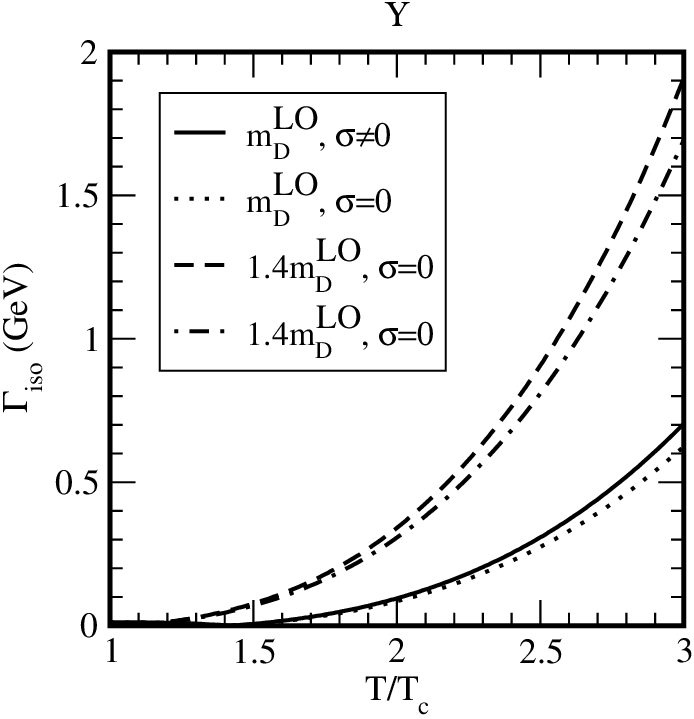}
\label{upsi}}
\end{subfigure}
\caption{\footnotesize Decay width of $ J/\psi $ (left) and $\Upsilon$
(right) states with and without nonperturbative (string) term in
an isotropic medium with the Debye masses in leading-order and
the lattice fitted result.} 
\label{gama1}
\end{figure}

The imaginary part of the potential, in small-distance limit, is a 
perturbation to the vacuum potential and thus provides an estimate 
for the width ($\Gamma$) for a resonance state and can be calculated,
in a first-order perturbation, by folding with the unperturbed 
(1S) Coulomb wavefunction 
\begin{eqnarray}
\Gamma_{\rm{(iso)}} =\left(\frac{4T}{\alpha m_Q^2}+
\frac{12\sigma T}{\alpha ^2m_Q^4}\right)~m_D^2~\log\frac{\alpha m_Q}{2m_D}.
\end{eqnarray}
%To see the effect of the nonperturbative term on the (thermal) width, we have 
%shown our results in Fig.(\ref{gama1}) for the ground
%states of $c\bar c$ and $b \bar b$ states, respectively. 
The main features of our results on the thermal width in (Fig.\ref{gama1}) are:
First the width always increases with the temperature.  Secondly the 
inclusion of the non vanishing nonperturbative string term, in addition
to the Coulomb term, makes the width larger than the earlier result with the 
perturbative Coulomb term~\cite{Dumitru11} only and thus the damping 
of the exchanged gluon in the heat bath provides larger contribution to the
dissociation rate and consequently reduce the yield of dileptons in the peak.
%%%%%%%%%%%%%%%%%%%%%%%%%%%Fig.3%%%%%%%%%%%%%%%%%%%%%%%%%
\vspace{0mm}
The effect of nonperturbative term on the width is relatively more on 
$J/\psi$ than
$\Upsilon$ state because the binding of $\Upsilon$ (1S) state is more
Coulombic than $J/\psi$ (1S) state. This may have far reaching implications
on the dissociation in medium.
Thirdly the width is also affected by the screening scale we chose to 
regulate the potential, {\em namely} the width with the higher 
screening scale (1.4 $m_D^{\rm{LO}}$) is more
than the leading-order result because the width, $\Gamma$ increases 
with the increase of the Debye mass. 

\subsubsection{Real and Imaginary Binding Energies: Dissociation
Temperatures}
To understand the in-medium properties of the quarkonium states,
one need to solve the Schr\"{o}dinger equation with both the real and imaginary 
part of the finite temperature potential. As seen earlier, in the 
short-distance limit, the vacuum contribution
dominates over the medium contribution whereas in the long-distance limit
the real part of the potential reduces to a Coulomb like 
potential and thus yields the real part of the binding energy in isotropic
medium:
\begin{eqnarray}
\rm{Re}~E_{\rm{bin}}^{\rm{iso}}\stackrel{\hat{r}\gg 1}{=}\left( \frac{m_Q\sigma^2 }{m_{{}_D}^4 n^{2}} +\alpha 
m_{{}_D} \right);~n=1,2 \cdot \cdot \cdot
\end{eqnarray}
However in the intermediate-distance ($rm_D \simeq 1$) scale, the
interaction becomes complicated and the potential
does not look simpler in contrast to the asymptotic limits, thus
the complex potential in general needs to be dealt with  
numerically to obtain the real and imaginary binding energies. 
There are some numerical methods to solve the Schr\"odinger equation
either in partial differential form (time-dependent) or eigen value form
(time-independent) by the
finite difference time domain method (FDTD) or matrix method, respectively.
In the later method, the stationary Schr\"odinger equation 
can be solved in a matrix form through a discrete basis, instead
of the continuous real-space position basis spanned by the states
$|\overrightarrow{x}\rangle$. Here the confining potential V is subdivided
into N discrete wells with potentials $V_{1},V_{2},...,V_{N+2}$ such that
for $i^{\rm{th}}$ boundary potential, $V=V_{i}$ for $x_{i-1} < x < x_{i};
~i=2, 3,...,(N+1)$. Therefore for the existence of a bound state, there
must be exponentially decaying wave function
in the region $x > x_{N+1}$ as $x \rightarrow  \infty $ and
has the form:
\begin{equation}
\Psi_{N+2}(x)=P_{{}_E} \exp[-\gamma_{{}_{N+2}}(x-x_{N+1})]+
Q_{{}_E} \exp [\gamma_{{}_{N+2}}(x-x_{N+1})] ,
\end{equation}
where, $P_{{}_E}= \frac{1}{2}(A_{N+2}- B_{N+2})$,
$Q_{{}_E}= \frac{1}{2}(A_{N+2}+ B_{N+2}) $ and,
$ \gamma_{{}_{N+2}} = \sqrt{2 \mu(V_{N+2}-E)}$. The eigenvalues
can be obtained by identifying the zeros of $Q_{E} $.
%%%%%%%%%%%%%%%%%%Figure 2 for potential%%%%%%%%%%%%%%%%%
\begin{figure}
\vspace{0mm}
\begin{subfigure}{
\includegraphics[width=5.9cm,height=6.5cm]{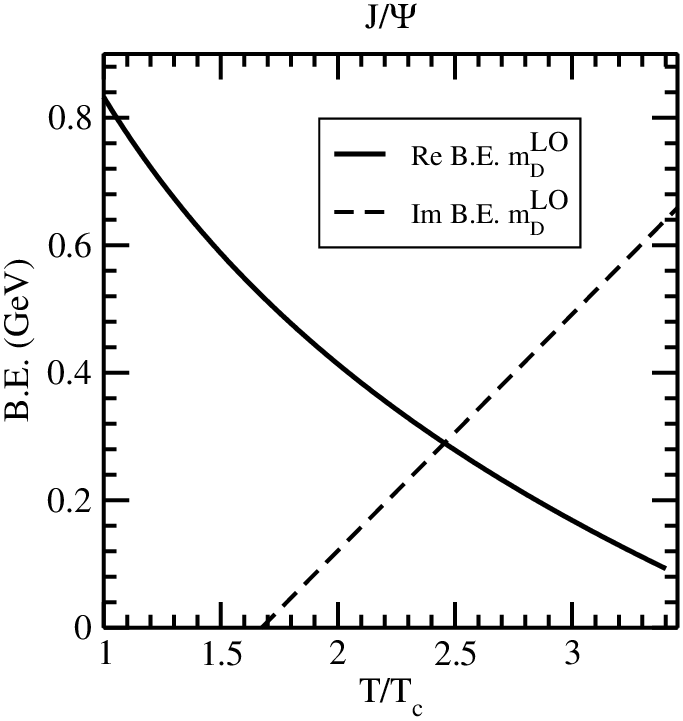} % Here is how to import EPS art
\label{jpsi}}
\end{subfigure}
\hspace{5mm}
\begin{subfigure}{
\includegraphics[width=5.9cm,height=6.5cm]{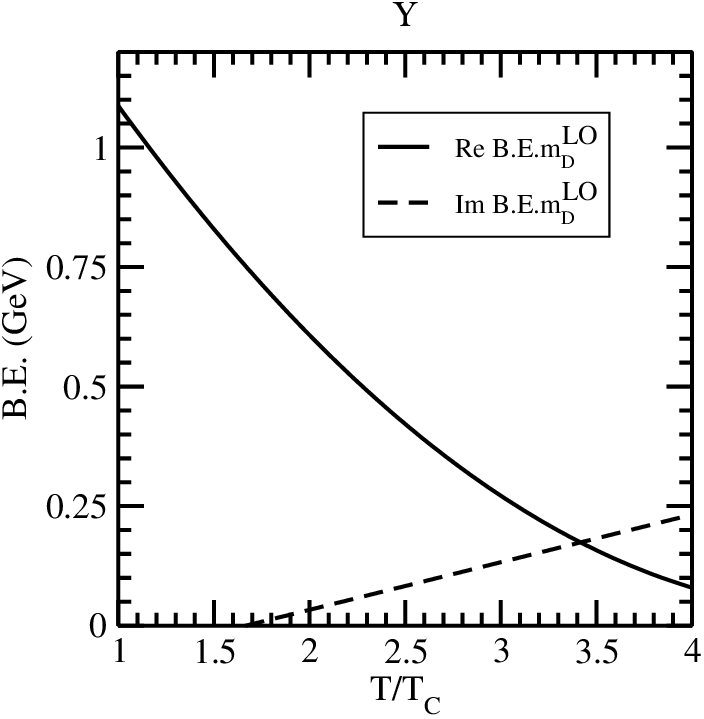}
\label{upsi}}
\end{subfigure}
\caption{\footnotesize Variation of the real and imaginary binding energies 
for the $J/\psi$ and $\Upsilon$ states with the 
temperature (in units of critical temperature, $T_c$) in the
left and right panel, respectively in an isotropic medium.}
\vspace{0mm}
\label{jpsi1}
\end{figure}

The binding energies shown in Fig. \ref{jpsi1} have the following features:
First when the nonperturbative term is included, the (real part) binding of 
$Q \bar Q$ pairs gets stronger with respect to the case where
only the Coulomb term is included. Secondly there is
a strong decreasing trend with the temperature because 
the screening becomes stronger with the increase of the temperature, so 
the real part of the potential becomes weaker compared to $T=0$ and
results in early dissolution of quarkonia in the medium.
Thirdly the real part of the binding energy decreases with the increase 
of screening scale (1.4 $m_D^{\rm{LO}}$). On the other hand the imaginary 
part of the binding energy increases with the temperature.
Thus the study of both the binding energies is poised to provide a 
wealth of information about the dissociation
pattern of quarkonium states in thermal medium which will
be used to determine the dissociation temperatures.

We will now study the dissociation in thermal medium 
to calculate the dissociation temperature ($T_{d}$) 
either from the intersection of the (real and imaginary) binding energies 
~\cite{Strickland:2011aa,Margotta:2011ta},
or from the conservative criterion on the width of the 
resonance as: $\Gamma \ge 2 \rm{Re}~{\rm{B.E.}}$ \cite{Mocsy07}. Although
both definitions are physically equivalent but they are numerically
different  (Table 1). For example, $J/\psi$ is dissociated
at 2.45 $T_c$ obtained from the intersection of binding energies while
the condition on width gives much lower temperature (1.40 $T_c$) .
Correspondingly $\Upsilon$ (1S) is dissociated at 3.40 $T_c$ and
3.10 $T_c$, respectively. Our results are found relatively higher 
compared to similar calculation~\cite{Strickland:2011aa,Margotta:2011ta},
which may be due to the absence of three-dimensional medium modification 
of the linear term in their calculation.

Finally we explore the sensitivity of the screening scale on the dissociation 
mechanism where the dissociation temperatures computed with the 
next-to-leading order (1.4 $m_D^{\rm{LO}}$) Debye mass are found
smaller than the leading-order result (Table 2). {\em For example}, $J/\psi$'s 
and $\Upsilon$'s are now dissociated at 1.33 $T_c$ and 1.91$T_c$, respectively.
%%%%%%%%%%%%%%%%%%%%%%%%%%%%%%%%%%%%%%%%%%%%%%%%%
\subsection{Potential in the anisotropic medium}
The space-time evolution of QGP relys on the viscous hydrodynamical 
treatment where the system assumes a local thermal equilibrium, i.e.
close to isotropic in momentum space, which may not be true
at the very early time in the collision of two nuclei, due to large 
momentum-space anisotropies~\cite{Martinez09,Martinez:2010sd,
Ryblewski:2010bs}. The degree of  anisotropy increases as the 
shear viscosity increases and thus one must address it while
calculating the heavy quark potential in the presence of momentum-space 
anisotropies. The real-part of the heavy quark potential was first 
considered in~\cite{Dumitru:2007hy} and then the imaginary part was 
obtained theoretically~\cite{Burnier:2009yu,Dumitru:2009fy, Philipsen:2009wg} 
as well as phenomenologically~\cite{Strickland:2011aa,Dumitru:2009ni,Margotta:2011ta}.
The main effect of the anisotropy is to reduce Debye screening which, in turn has the effect 
that heavy quarkonium states can survive upto higher temperatures.
However the aforesaid works in anisotropic medium are limited to the 
medium-modification of the 
perturbative part only and the nonperturbative string term was assumed to zero.
However, the string-tension is non vanishing even at temperatures much beyond 
the deconfinement point~\cite{string1,string2,string3}, so
one should study its effect on the heavy quark potential 
in anisotropic medium too.
%%%%%%%%%%%%%%%%%%%%%%%%%%%%%%%%%%%%%
\subsubsection{Real Part of the Potential}
\begin{figure}
%\centering
\begin{subfigure}{
\includegraphics[width=5.9cm,height=6.5cm]{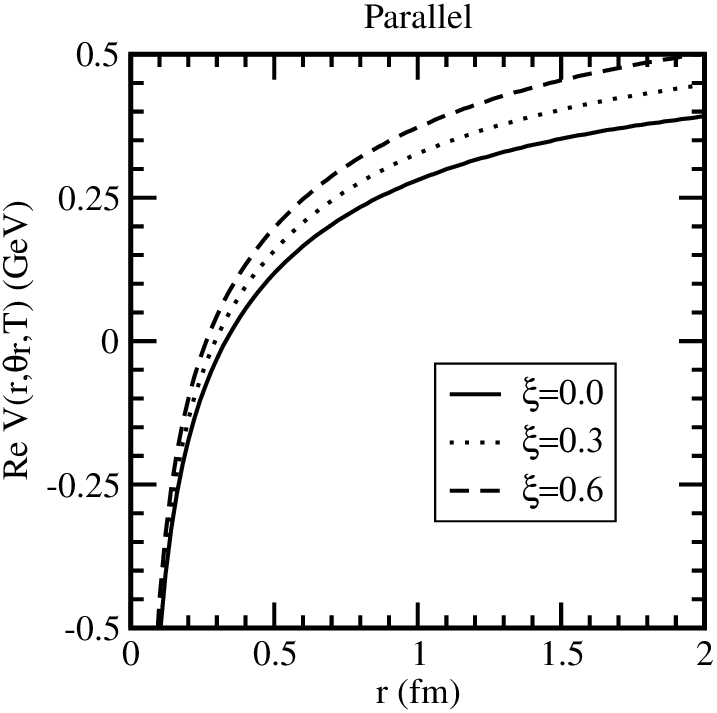}
\label{para_plot_a}}
\end{subfigure}
 \hspace{5mm}
\begin{subfigure}{
 \includegraphics[width=5.9cm,height=6.5cm]{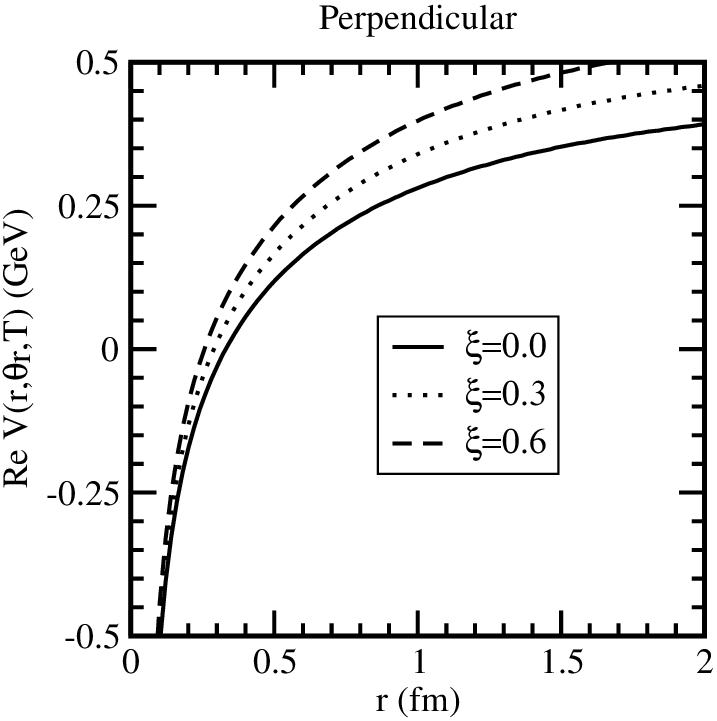}
\label{para_plot_b}}
\end{subfigure}
\caption{\footnotesize Real part of the potential for both the parallel
(left)and perpendicular (right) alignment with the Debye mass 
in the leading-order.}
\vspace{2cm}
\label{para_plot11}
\end{figure}

Like in isotropic medium, we obtain the real-part of the potential 
in weakly-anisotropic medium~\cite{lt} from the anisotropic
corrections to the (temporal component) real-part of retarded propagator
(\ref{rtrdprop})
\begin{eqnarray}
\label{pot}
Re V_{\rm(aniso)}({\bf r},\xi,T)&=&\int \frac{d^3\mathbf p}{{(2\pi)}^{3/2}}
 (e^{i\mathbf{p} \cdot \mathbf{r}}-1)
 \left(-\sqrt{(2/\pi)}\frac{\alpha}{p^2}- 
\frac{4\sigma}{\sqrt{2 \pi} p^4}\right) \times \nonumber\\
&&p^2\left[\frac{1}{(p^2+m_D^2)}-\frac{\xi m_D^2}{6(p^2+m_D^2)^2}
(3\cos(2\theta_p)-1)\right] \nonumber\\
&\equiv& Re V_{1(aniso)}({\bf r},\xi,T)+ Re V_{2(aniso)}({\bf r},\xi,T)~,
\end{eqnarray}
%where $\theta_p$ is the angle between ${\bf r}$ and ${\bf n}$
where $Re V_{1(aniso)} ({\bf r},\xi,T)$ and $Re V_{2(aniso)}(\mathbf{r},\xi,T)$ are the 
medium modifications corresponding to the Coulomb and 
string term, respectively, are given by
\begin{eqnarray}
Re V_{1(aniso)}({\bf r},\xi,T) &=&-\frac{\alpha}{2\pi^{2}}\int d^{3}\mathbf{p} 
(e^{i\mathbf{p} \cdot \mathbf{r}}-1)\left[\frac{1}{(p^2+m_D^2)}
-\frac{\xi m_D^2}{6(p^2+m_D^2)^2}
(3\cos 2\theta_p-1)\right]\\
Re V_{2(aniso)}({\bf r},\xi,T) &=&-\frac{4\sigma}{{2\pi}^{2}}\int d^{3}\mathbf{p} 
(e^{i\mathbf{p} \cdot \mathbf{r}}-1)\frac{1}{p^2}\left[\frac{1}{(p^2+m_D^2)}
-\frac{\xi m_D^2}{6(p^2+m_D^2)^2}
(3\cos 2\theta_p-1)\right]~.
\end{eqnarray}
To perform the momentum integration, we use the transformation 
$\cos\theta_p = \cos\theta_r \cos\theta_{pr} +
\sin\theta_r \sin\theta_{pr} \cos\phi_{pr}$,
where $ \theta_p $ and $ \theta_r $ are the angles between {\bf p} and 
{\bf n}, {\bf r} and {\bf n}, respectively and 
$ \theta_{pr} $, $ \phi_{pr} $ are the angular variables between the
vectors, ${\bf p}$ and ${\bf r}$. So after the integration, the Coulombic 
contribution to the potential becomes
\begin{eqnarray}
Re V_{1(aniso)}({\bf r},\xi,T) &=&-\alpha m_D\left[\frac{e^{-\hat{r}}}{\hat{r}}+1
+\xi\left[\frac{(e^{-\hat{r}}-1)}{6}
+\left[e^{-\hat{r}}\left(\frac{1}{6}+\frac{1}{2\hat{r}}
+\frac{1}{\hat{r}^2}\right)
+\frac{(e^{-\hat{r}}-1)}{\hat{r}^3}\right]\right.\right. \times \nonumber\\
&& \left. \left.(1-3\cos^2\theta_r)\right]\right]~,
\label{col}
\end{eqnarray}
and the string contribution is
\begin{eqnarray}
Re V_{2(aniso)}({\bf r},\xi,T) &=&\frac{2\sigma}{m_{{}_D}}
\left[\frac{(e^{-\hat{r}}-1)}{\hat{r}}+1
+2\xi\left[\left(\frac{(e^{-\hat{r}}-1)}{6\hat{r}}+\frac{e^{-\hat{r}}}{12}+\frac{1}{6}\right)
\right.\right.\nonumber\\
&& \left.\left.+\left(e^{-\hat{r}}\left(\frac{1}{{\hat{r}}^2}+\frac{5}{12\hat{r}}+\frac{1}{12}\right)+\frac{1}{12\hat{r}}+\frac{(e^{-\hat{r}}-1)}{{\hat{r}}^3}\right)(1-3\cos^2\theta_r)\right]\right]~.
\label{strng}
\end{eqnarray}
Thus the real-part of the potential in the anisotropic medium becomes 
\begin{eqnarray}
Re V_{aniso}(r,\theta_r,T) &=& \left(\frac{2\sigma}{m_{{}_D} }-\alpha m_{{}_D} \right) 
\frac{e^{-\hat r}}{\hat{r}}-\frac{2\sigma}{m_{{}_D}\hat{r}}  
+\frac{2\sigma}{m_{{}_D}}- \alpha m_{{}_D} \nonumber\\
&+&  \xi \left( \frac{2 \sigma} {m_{D}}\frac{e^{-\hat{r}}}{\hat{r}} 
\left[\frac{e^{\hat{r}}-1}{\hat{r}^2}-\frac{5 e^{\hat{r}}}{12}+\frac{\hat{r}e^{\hat{r}}}{3}-
\frac{1}{\hat{r}}+\frac{\hat{r}}{12} -\frac{1}{12}\right] \right. \nonumber\\
&-& \left. \frac{\alpha m_{D}} {2}\frac{e^{-\hat{r}}}{\hat{r}}
\left[\frac{e^{\hat{r}}-1}{\hat{r}^2}-\frac{1} {\hat{r}}-\frac{\hat{r}e^{\hat{r}}}{3}+
\frac{\hat{r}}{6}-\frac{1}{2}\right] \right. \nonumber\\
&+ & \left( \left. \frac{2 \sigma} {m_{D}}\frac{e^{-\hat{r}}}{\hat{r}}\left[
3 \frac{e^{\hat{r}}-1}{\hat{r}^2}-\frac{e^{\hat{r}}}{4}-
\frac{3}{\hat{r}}-\frac{\hat{r}}{4}-\frac{5}{4}\right] \right.\right.\nonumber\\
&-&\frac{\alpha m_{D}}{2} \frac{e^{-\hat{r}}}{\hat{r}}\left[ \left.\left.
3 \frac{e^{\hat{r}}-1}{\hat{r}^2}-\frac{3}{\hat{r}}-\frac{\hat{r}}{2}-
\frac{3}{2}\right] \right) \cos~2 \theta_r \right) \nonumber\\
&=& Re V_{iso}(r,T) + V_{\rm{tensor}}(r,\theta_r,T). 
\label{fullp}
\end{eqnarray}
Thus the anisotropy in the momentum space introduces an angular ($\theta_r$)
dependence, in addition to the interparticle separation ($r$), to the 
real part of the potential, in contrast to the $r$-dependence only in 
an isotropic medium. The real potential becomes stronger with the 
increase of anisotropy (shown in Fig.\ref{para_plot11})
because the (anisotropic) Debye mass $m_D(\xi,T)$ 
(or equivalently angular-dependent Debye mass $m_D(\theta_r,T)$)
in an anisotropic
medium is always smaller than in an isotropic medium. As a result
the screening of the Coulomb and string contribution are
less accentuated, compared to the isotropic medium.
In particular the potential for quark pairs aligned in the 
direction of anisotropy are stronger than the pairs aligned in the 
transverse direction.

%%%%%%%%%%%%%%%%%%Figure 4 for potential%%%%%%%%%%%%%%%%%

%%%%%%%%%%%%%%%%%%%%%%%%%%%%%%%%%%%%%%%%%%%%%%%%%%%%%%%%%%%%%%%%%%%%%
\subsubsection{Imaginary Part of the Potential: Thermal width, 
$\Gamma_{\rm{aniso}}$}
Recently the imaginary part with a momentum-space anisotropy and its effects
on the thermal widths of the resonance states have been 
studied~\cite{Margotta:2011ta,Dumitru09,Dumitru11,Guo09}, with the 
medium-modification to the perturbative (Coulomb) term only.
The imaginary part of the potential arises due to the singlet-to-octet 
transitions induced by the dipole vertex as well as due to the 
Landau damping in the plasma, {\em i.e.} scattering of the gluons with 
space-like momentum off the thermal excitations in the medium. We follow 
their work by including the medium corrections to both 
perturbative (Coulombic) and  non-perturbative (string) terms
in a weakly anisotropic medium. Like in isotropic medium, we can obtain 
the imaginary part of the potential by the leading anisotropic
correction to the imaginary part of the (temporal component) symmetric 
propagator as
\begin{eqnarray}
Im V_{\rm(aniso)}({\bf r},\xi,T)&=&-\int \frac{d^3\mathbf{p}}{(2\pi)^{3/2}}
(e^{i\mathbf{p} \cdot \mathbf{r}}-1)
\left(-\sqrt{\frac{2}{\pi}}\frac{\alpha}{p^2}-\frac{4\sigma}{\sqrt{2\pi}p^4}\right)
p^2\left[\frac{-\pi T m_D^2}{p(p^2+m_D^2)^2}\right.\nonumber\\
&& \left.+\xi[\frac{3\pi T m_D^2}{2p(p^2+m_D^2)^2}\sin^2{\theta_p}
-\frac{4\pi T m_D^4}{p(p^2+m_D^2)^3}(\sin^2\theta_p-\frac{1}{3})\right]
\nonumber\\
&& \equiv Im V_{1(aniso)} ({\bf r},\xi,T)+ Im V_{2(aniso)} ({\bf r},\xi,T) ~,
\end{eqnarray}
where $Im V_{1(aniso)} ({\bf r},\xi,T)$ and 
$Im V_{2(aniso)} (\mathbf{r},\xi,T)$ are the
imaginary contributions corresponding to 
the Coulombic and linear terms in anisotropic medium, respectively:
\begin{eqnarray}
Im V_{1\rm(aniso)}({\bf r},\xi,T)&=&\frac{\alpha}{2\pi^{2}}\int d^{3}\mathbf{p} 
(e^{i\mathbf{p} \cdot \mathbf{r}}-1)
\left[\frac{-\pi T m_D^2}{p(p^2+m_D^2)^2}+\xi\left[\frac{3\pi T m_D^2}{4p(p^2+m_D^2)^2}\sin^2{\theta_p}\right.\right.\nonumber\\
&&\left. \left.-\frac{2\pi T m_D^4}{p(p^2+m_D^2)^3}(\sin^2\theta_p-
\frac{1}{3})\right] \right],\\
Im V_{2\rm(aniso)}({\bf r},\xi,T)&=&\frac{4\sigma}{({2\pi})^2}\int \frac{d^3\mathbf{p}}
{(2\pi)^{3/2}}(e^{i\mathbf{p} \cdot \mathbf{r}}-1)
\frac{1}{p^2} \left[\frac{-\pi T m_D^2}{p(p^2+m_D^2)^2}
+\xi\left[\frac{3\pi T m_D^2}{4p(p^2+m_D^2)^2}\sin^2{\theta_p}\right.\right.
\nonumber\\~
&&\left. \left. -\frac{2\pi T m_D^4}{p(p^2+m_D^2)^3}
(\sin^2\theta_p-\frac{1}{3})\right]\right].
\end{eqnarray}
%%%%%%%%%%%%%%%%%%Figure 6%%%%%%%%%%%%%%%%%%%%%%%%
\begin{figure}
\vspace{0mm}
\begin{subfigure}{
\includegraphics[width=5.9cm,height=6.5cm]{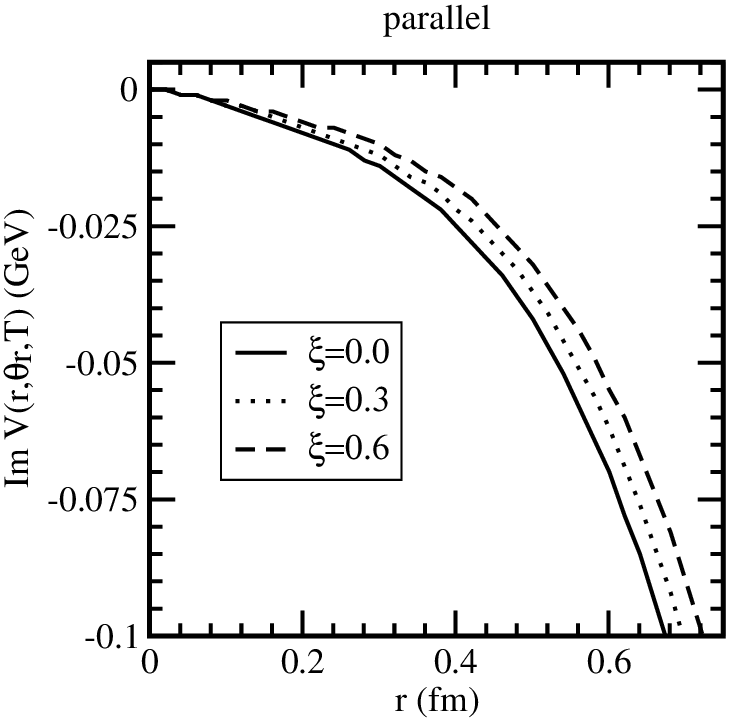} % Here is how to import EPS art
\label{jpsi}}
\end{subfigure}
\hspace{5mm}
\begin{subfigure}{
\includegraphics[width=5.9cm,height=6.5cm]{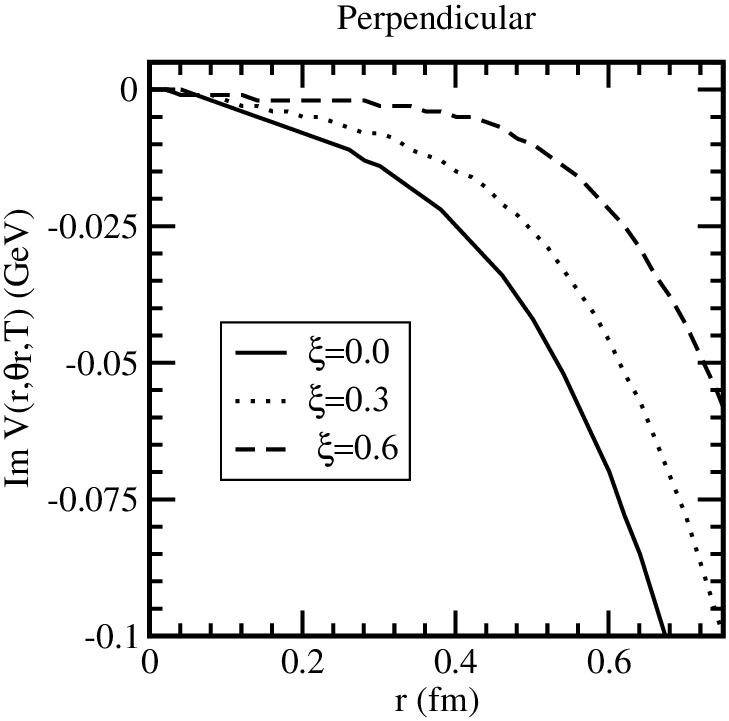}
\label{upsi}}
\end{subfigure}
\caption{\footnotesize Imaginary part of the potential for parallel (left)
and perpendicular (Right) alignment in an anisotropic medium.}
\vspace{0mm}
\label{upsi1}
\end{figure}
Since the isotropic contribution is already calculated 
in Sec.2.2.2, so the anisotropic contribution to the
perturbative term in the leading-order is given by~\cite{Dumitru09}
\begin{eqnarray}
Im V_{1(aniso)}({\bf r},\xi,T)&\equiv & \alpha T \xi\left[\phi_{1}(\hat{r}, \theta_r)
+\phi_2(\hat{r},\theta_r)\right],                                     
\end{eqnarray}
where the functions $\phi_{1}(\hat{r},\theta_r)$ and 
$\phi_{2}(\hat{r},\theta_r)$ are
\begin{eqnarray}
\phi_{1}(\hat{r},\theta_r)&=& \frac{{\hat{r}}^2}{600} \left[123-
90 \gamma_{E}- 90\log\hat{r}     
+\cos(2\theta_r) \left(-31+30\gamma_{E}+30\log\hat{r}\right)\right], \nonumber\\  
\phi_{2}(\hat{r},\theta_r)&=& \frac{{\hat{r}}^2}{90}(-4+ 3 
\cos(2\theta_r)).                
\end{eqnarray}
Similarly the imaginary contribution due to the nonperturbative (linear) term 
can also separated into the isotropic and anisotropic term, where
the isotropic part is already calculated in Sec.2.2.2 and hence the 
anisotropic part is now calculated 
\begin{eqnarray}
Im V_{2(aniso)}(r,\xi,T)=-\xi~\frac{2\sigma T}{m_D^2}
\left[\psi_1(\hat{r},\theta_r)+\psi_2(\hat{r},\theta_r)\right]~.
\label{v2aniso}
\end{eqnarray}
The function, $\psi_1 (\hat{r},\theta_r)$ 
is given by 
\begin{eqnarray}
\psi_1(\hat{r},\theta_r)&=&\int \frac{dz}{z(z^2+1)^2}\left[1-\frac{3}{2}
\left(\sin^2\theta_r\frac {\sin{z\hat r}}{z\hat r}
+(1-3\cos^2\theta_r)G(\hat{r},z)\right)\right]~,
\end{eqnarray}
where $G(\hat{r},z)$ is given by
\begin{eqnarray}
G(\hat{r},z)&=&\frac{z\hat r\cos(z\hat r)-\sin(z\hat r)}{(z\hat r)^3}.
\end{eqnarray}
Substituting $G(\hat{r},z)$ into $\psi_1(\hat{r},\theta_r)$ and
decomposing into $\theta_r$- dependent and independent terms,
the function, $\psi_1(\hat{r},\theta_r)$ can be rewritten as
\begin{eqnarray}
\psi_1(\hat{r},\theta_r)&=&\int \frac{dz}{z(z^2+1)^2}\left[1-\frac{3}{2}\left(
\frac {\sin(z\hat r)}{z\hat r}+\frac{\cos(z\hat r)}{(z\hat r)^2}-\frac{\sin(z\hat r)}{(z\hat r)^3}\right)\right.\nonumber\\
&& \left.+\frac{3}{2}\left(\frac{\sin(z\hat r)}{z\hat r}+3\frac{\cos(z\hat r)}{(z\hat r)^2}
-3\frac{\sin(z\hat r)}{(z\hat r)^3}\right)\cos^2\theta_r)\right]\nonumber\\
&\equiv&\psi_1^{(1)}(\hat{r})+\psi_1^{(2)}(\hat{r},\theta_r) ~,
\label{psi_1}
\end{eqnarray}
where the functions $ \psi_1^{(1)}(\hat{r}) $ and 
$ \psi_1^{(2)} (\hat{r},\theta_r) $ are given by
\begin{eqnarray}
\psi_1^{(1)}(\hat{r})&=&\int \frac{dz}{z(z^2+1)^2}\left[1-\frac{3}{2}\left(
\frac {\sin(z\hat r)}{z\hat r}+\frac{\cos(z\hat r)}{(z\hat r)^2}-\frac{\sin(z\hat r)}{(z\hat r)^3}\right)\right]\nonumber\\
&=&\hat r^4\int \frac{dx}{x(x^2+\hat r^2)^2}\left[1-\frac{3}{2}\left(
\frac {\sin(x)}{x}+\frac{\cos(x)}{x^2}-\frac{\sin(x)}{x^3}\right)\right]\nonumber\\
&=&\frac{\hat r^2}{10}+\frac{(-739+420\gamma_E +420\log(\hat r))\hat r^4}{39200}+O(\hat r^5),
\end{eqnarray}
and
\begin{eqnarray}
\psi_1^{(2)}(\hat{r},\theta_r)&=&\frac{3}{2}\int \frac{dz}{z(z^2+1)^2}\left[\left(\frac{\sin(z\hat r)}{z\hat r}+3\frac{\cos(z\hat r)}{(z\hat r)^2}
-3\frac{\sin(z\hat r)}{(z\hat r)^3}\right)\cos^2\theta_r)\right]\nonumber\\
&=&\frac{3}{2}\hat r^4\int \frac{dx}{x(x^2+\hat r^2)^2}\left[\left(\frac{\sin(x)}{x}+\frac{3\cos(x)}{x^2}
-\frac{3\sin(x)}{x^3}\right)\cos^2\theta_r \right]\nonumber\\
&=&\left(-\frac{r^2}{20}+\frac{(176-105\gamma_E -105\log(\hat r))\hat r^4}{14700}+O(\hat r^5)\right)\cos^2\theta_r~.
\end{eqnarray}
The remaining function in the imaginary part of the potential associated 
with the linear term (\ref{v2aniso}) can similarly be separated into 
$\theta_r$- dependent and independent terms:
\begin{eqnarray}
\psi_2(\hat{r},\theta_r)&=&-\frac{4}{3}\int\frac{dz}{z(z^2+1)^3}
\left[1-3\left[(\frac{2}{3}-\cos^2\theta_r)\frac {\sin{z\hat r}}{z\hat r}
+(1-3\cos^2\theta_r)G(\hat{r},z)\right]\right]\nonumber\\
&=&-\frac{4}{3}\int\frac{dz}{z(z^2+1)^3}\left[ \left(1-
\frac {2\sin{z\hat r}}{z\hat r}-\frac{3\cos(z\hat r)}{(z\hat r)^2}+\frac{3\sin(z\hat r)}{(z\hat r)^3}\right)\right.\nonumber\\
&& \left.+ 3\left(\frac {\sin{z \hat r}}{z\hat r}+\frac{3\cos(z\hat r)}{(z\hat r)^2}-\frac{3\sin(z\hat r)}{(z\hat r)^3}\right)\cos^2\theta_r \right]\nonumber\\
&\equiv &\psi_2^{(1)}(\hat{r})+\psi_2^{(2)}(\hat{r},\theta_r),
\label{psi_2}
\end{eqnarray}
where the functions $ \psi_2^{(1)}(\hat{r}) $ and 
$ \psi_2^{(2)}(\hat{r},\theta_r) $  are given by
\begin{eqnarray}
\psi_2^{(1)}(\hat{r})&=&-\frac{4}{3}\int\frac{dz}{z(z^2+1)^3} \left(1-
\frac {2\sin{z\hat r}}{z\hat r}-\frac{3\cos(z\hat r)}{(z\hat r)^2}+\frac{3\sin(z\hat r)}{(z\hat r)^3}\right)\nonumber\\
&=&-\frac{4}{3}\left[\frac{7 \hat r^2}{120}-\frac{11 \hat r^4}{3360}+O(\hat r^5)\right],
\end{eqnarray}
and
\begin{eqnarray}
\psi_2^{(2)}(\hat{r},\theta_r)&=&-4\int\frac{dz}{z(z^2+1)^3} \left(\frac {\sin{z\hat r}}{z\hat r}+\frac{3\cos(z\hat r)}{(z\hat r)^2}-\frac{3\sin(z\hat r)}{(z\hat r)^3}\right)\cos^2\theta_r \nonumber\\
&=&-4\left[-\frac{\hat r^2}{60}+\frac{\hat r^4}{840}+O(\hat r^5)\right]\cos^2\theta_r .
\end{eqnarray}
So the functions $ \psi_1(\hat{r},\theta_r) $ and $ \psi_2(\hat{r},\theta_r)$
are finally given by
\begin{eqnarray}
\psi_1(\hat{r},\theta_r)&=&\frac{\hat{r}^2}{10}+\frac{(-739+420\gamma_E 
+420\log(\hat r))\hat r^4}{39200}\nonumber\\
&+&\left(-\frac{\hat r^2}{20}
+\frac{(176-105\gamma_E -105\log(\hat r))\hat r^4}{14700}\right)
\cos^2\theta_r,\\
 \psi_2(\hat{r},\theta_r)&=&-\frac{4}{3}\left[\frac{7 \hat r^2}{120}-
\frac{11 \hat r^4}{3360}+O(\hat r^5)\right] \nonumber\\
 &&-4\left[-\frac{\hat r^2}{60}+\frac{\hat r^4}{840}+O(\hat r^5)\right]
\cos^2\theta_r~,
\end{eqnarray}
respectively and $\gamma_E  $ is the Euler-Gamma constant.
%%%%%%%%%%%%%%%%%%%%%%%%Figure 7%%%%%%%%%%%%%%%%%%%%%%%%
\begin{figure}
\vspace{0mm}
\begin{subfigure}{
\includegraphics[width=5.9cm,height=6.5cm]{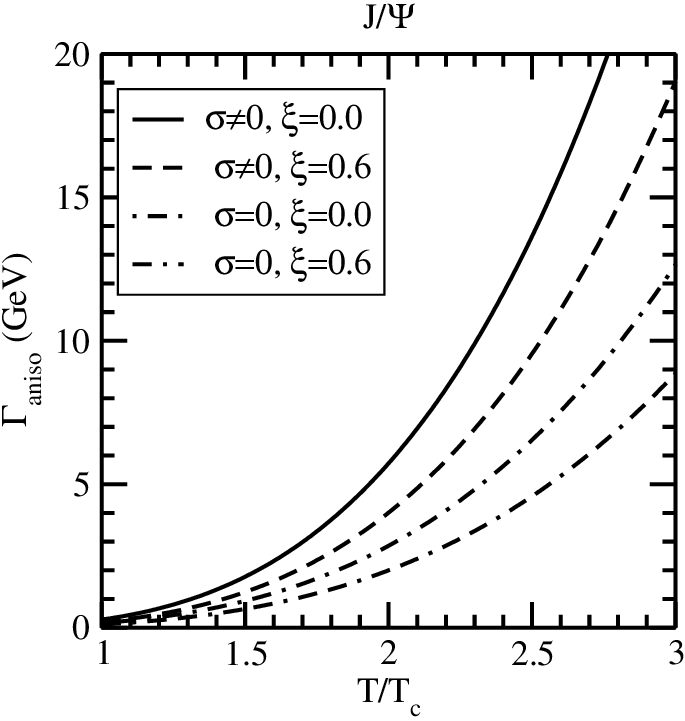}} % Here is how to import EPS art
%\label{jpsi}}
\end{subfigure}
\hspace{5mm}
\begin{subfigure}{
\includegraphics[width=5.9cm,height=6.5cm]{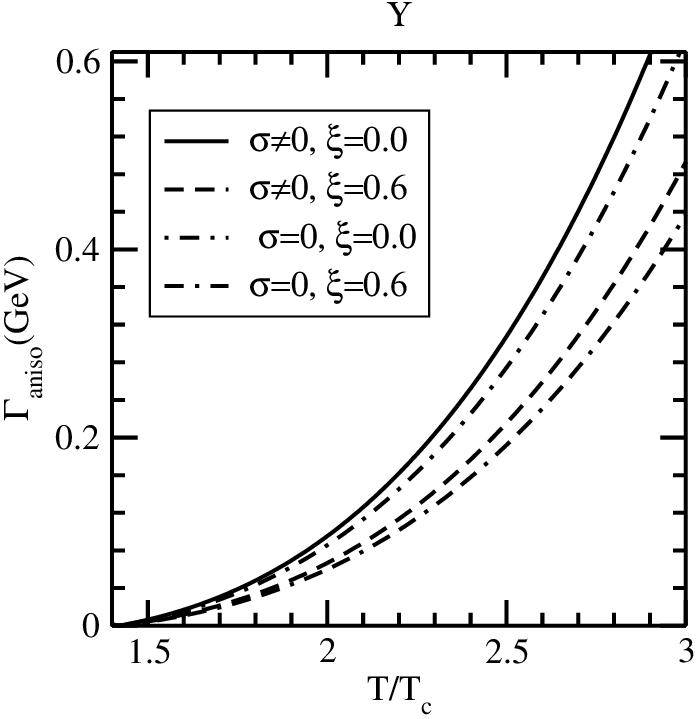}}
%\label{upsi}}
\end{subfigure}
\caption{\footnotesize The thermal width for the $J/\psi$ and $\Upsilon$ 
states in the anisotropic medium.}
\label{gamaniso}
\end{figure}
\vspace{0mm}
%%%%%%%%%%%%%%%%%%%%%%%%%%%%%%%%%%%%
Finally the short and long-distance contributions, in the leading logarithmic
 order 
\begin{eqnarray}
Im V_{1(aniso)} (r,\theta_r,T)&=&-\alpha T{\hat r^2}\log(\frac{1}{\hat r})
\left(\frac{1}{3}-\xi\frac{3-\cos2\theta_r}{20}\right),\\
Im V_{2(aniso)} (r,\theta_r,T)&=&-\frac{2\sigma T}{m_D^2}\frac{\hat r^4}{20}\log(\frac{1}{\hat r})
\left(\frac{1}{3}-\xi\frac{2-\cos2\theta_r}{14}\right)~,
\end{eqnarray}
gives the imaginary part of the potential in the anisotropic medium
\begin{eqnarray}
Im V_{\rm{(aniso)}} (r,\theta_r,T)&=&-T\left(\frac{\alpha {\hat r^2}}{3}
+\frac{\sigma {\hat r}^4}{30m_D^2}\right)\log(\frac{1}{\hat r})\nonumber\\
&&+\xi T\left[\left(\frac{\alpha {\hat r^2}}{5}+\frac{3\sigma {\hat r^4}}{140m_D^2}\right)\right.
\left.-\cos^{2}\theta_r \left(\frac{\alpha {\hat r^2}}{10}+\frac{\sigma {\hat r^4}}{70m_D^2}\right)\right]\log(\frac{1}{\hat r}) ~,
\end{eqnarray}
which is found to be smaller than the isotropic medium and decreases with the 
increase of anisotropy (shown in Fig.\ref{upsi1}).
%%%%%%%%%%%%%%%%%TABLE 1 oneloop%%%%%%%%%
\begin{table}
\centering
\begin{tabular}{|c|c|c|c|c|}
\hline
Method  & State  &$\xi=0.0$ & $\xi=0.3$ & $\xi=0.6$\\
\hline\hline
 Re B.E.=Im B.E.& $\jpsi $ & 2.45& 2.46& 2.47 \\
 & $\Upsilon $ & 3.40& 3.45& 3.46 \\
   \hline
$\Gamma$=2B.E. & $\jpsi  $ & 1.40& 1.46& 1.54 \\
 & $\Upsilon$ & 3.10& 3.17&3.26 \\
   \hline
\end{tabular}
\caption {\footnotesize {Dissociation temperatures of  $J/\psi$ 
and $\Upsilon$ states for different anisotropies with the
Debye mass in leading-order.}}
\label{tdlo}
\end{table}

%%%%%%%%%%%%%%%%%TABLE 2 oneloop%%%%%%%%%
\begin{table}
\centering
\begin{tabular}{|c|c|c|c|c|}
\hline
Method  & State  &$\xi=0.0$ & $\xi=0.3$ & $\xi=0.6$\\
\hline\hline
 Re B.E.=Im B.E.& $\jpsi $ & 1.33& 1.34& 1.35 \\
 & $\Upsilon $ &1.91& 1.93& 1.94 \\
   \hline
$\Gamma$=2B.E. & $\jpsi  $ & 1.02& 1.06& 1.12 \\
 & $\Upsilon$ &1.88& 1.92&2.02 \\
   \hline
\end{tabular}
\caption {\footnotesize {The same as Table 1, having the
Debye mass ($m_D=1.4~m_D^{\rm{LO}}$)}}
\label{tdlat}
\end{table}
Like in isotropic medium, in weakly anisotropic medium too, the imaginary 
part is found to be a perturbation and thus provides 
an estimate for the (thermal) width for a particular resonance 
state: 
\begin{eqnarray}
\Gamma_{\rm(aniso)} &=& \int d^3 {\bf r}|\Psi(r)|^2\left[\alpha T{\hat r^2}
\log(\frac{1}{\hat r})\left(\frac{1}{3}-\xi
\frac{3-\cos 2\theta_r}{20}\right)\right.\nonumber\\
&&\left.+\frac{2\sigma T}{m_D^2}{\hat r^4}\log(\frac{1}{\hat r})
\frac{1}{20}\left(\frac{1}{3}-\xi\frac{2-\cos2\theta_r}{14}\right)\right]\nonumber\\
&=&T\left(\frac{4}{\alpha m_Q^2}+\frac{12\sigma}{\alpha ^2m_Q^4}\right) 
\left(1-\frac{\xi}{2}\right)m_D^2 \log\frac{\alpha m_Q}{2m_D}~,
\end{eqnarray}
which shows that the width in anisotropic medium becomes smaller than
in isotropic medium and gets narrower with the increase of anisotropy
(shown in Fig. \ref{gamaniso}).
This is due to the fact that $\Gamma$ is approximately proportional to the
(square) Debye mass and the Debye mass decreases in the anisotropic medium because the 
effective local parton density around a test (heavy) quark is smaller 
compared to isotropic medium.

%%%%%%%%%%%%%%%%%%%%%%%%%%%%%%%%%%%%%%%%%%%%%%%%%%%%%%%%%%%%%%%%%%%%%
\subsubsection{Real and Imaginary Binding Energies: Dissociation temperatures}
The real part of the potential thus obtained in anisotropic 
medium (\ref{fullp}), in contrast to its counterpart (spherically 
symmetric potential) in isotropic medium eq. (\ref{pis}) is non-spherical and so one 
cannot simply obtain the energy eigen values
by solving the radial part of the Schr\"{o}dinger equation alone because
the radial part is no longer sufficient due to
the angular dependence in the potential. Other way to understand
is that because of the anisotropic
screening scale, the wave functions are no longer radially symmetric for
$\xi \ne 0$.  So one has to solve the potential in three dimension
but in the small $\xi$-limit, the non-symmetric component
$V_{\rm{tensor}}(r,\theta_r,T)$ is much smaller than the
symmetric (isotropic) component $Re V_{\rm(iso)}(r,T)$ and thus
can be treated as perturbation.
%%%%%%%%%%%%%%%%%%Figure 5%%%%%%%%%%%%%%%%%%%%%%%%
\begin{figure*}
\vspace{2cm}
\begin{subfigure}{
\includegraphics[width=5.9cm,height=6.5cm]{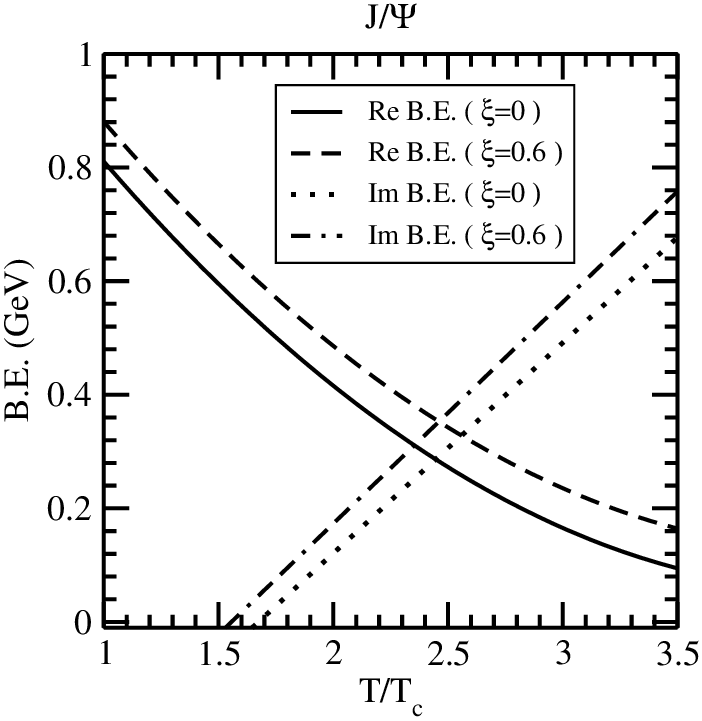} % Here is how to import EPS art
%\label{jpsi}
}
\end{subfigure}
\hspace{5mm}
\begin{subfigure}{
\includegraphics[width=5.9cm,height=6.5cm]{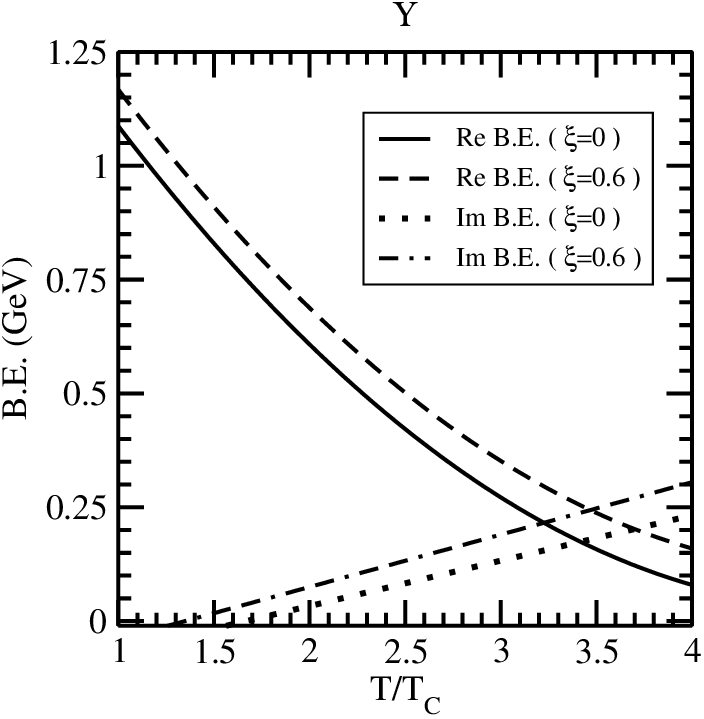}
%\label{upsi}
}
\end{subfigure}
\caption{\footnotesize Variation of the real and imaginary part of the 
binding energies for $J/\psi$ and $\Upsilon$ states for different 
anisotropies.}
\vspace{0mm}
\label{ju1}
\end{figure*}
Therefore, the corrected energy eigen values come from the solution of 
Schr\"odinger equation of the isotropic component 
plus the first-order perturbation due to the anisotropic
component $V_{\rm{tensor}}(r,\theta_r,T)$. 

In short-distance limit, the vacuum contribution dominates over the 
medium contribution even for the weakly anisotropic medium and in the 
long-distance limit, the real part of potential in high temperature
approximation results a Coulomb plus a subleading anisotropic contribution :
\begin{eqnarray}
\label{largp}
Re V_{\rm{(aniso)}}(r,\theta_r,T) &\stackrel{\hat{r}\gg1}{\simeq}& -\frac{2\sigma}{m^2_{{}_D}r}
-\alpha m_{{}_D} -\frac{5\xi}{12}~\frac{2\sigma}{m^2_{{}_D}r} 
\left(1+\frac{3}{5}\cos 2\theta_r \right) \\
&\equiv & Re V_{\rm{iso}} (\hat{r} \gg 1,T)+ V_{\rm{tensor}} (\hat{r} \gg 1~.
\theta_r,T)~.
\end{eqnarray}
where the anisotropic contribution
($V_{\rm{tensor}} (\hat{r} \gg 1, \theta_r,T)$) is smaller than the isotropic
one ($Re V_{\rm{iso}} (\hat{r} \gg 1,T)$), so the anisotropic part
can be treated as perturbation. Therefore, the real part of binding energy
may be obtained from the radial part of the Schr\"odinger equation
(of the isotropic component) plus the first-order perturbation
due to the anisotropic component as 
\begin{eqnarray}
{\rm Re}~E_{\rm{bin}}^{\rm{aniso}} =\left( \frac{m_Q\sigma^2 }{m_{{}_D}^4 n^{2}} +
\alpha m_{{}_D} \right) +
\frac{2\xi}{3} \frac{m_Q\sigma^2 }{m_{{}_D}^4 n^{2}} ,
\end{eqnarray}
where the first term is the solution of (radial-part) of
the Schr\"odinger equation with the isotropic part 
($Re V_{\rm{iso}} (\hat{r} \gg 1,T)$) 
and the second term is due to the anisotropic perturbation
of the  tensorial component ($V_{\rm{tensor}} (\hat{r} \gg 1, \theta_r,T)$)
calculated from the first-order perturbation theory.

%The real part binding energies for $J/\psi$ and $\Upsilon$ state are computed 
The real and imaginary part of the binding energies for the $J/\psi$ 
and $\Upsilon$ states are computed 
numerically in Fig.~\ref{ju1} for different values of anisotropies, with
the following observations: The inclusion of the nonperturbative 
string term makes the quarkonium states more bound in the anisotropic medium
too. Secondly the binding of $Q \bar Q$ pairs becomes stronger with respect to 
their isotropic counterpart and increases with the increase of anisotropy
because the (real part) potential becomes deeper due to the weaker 
screening. Last but not the least, as the screening scale increases 
the binding gets weakened even in the anisotropic medium. In contrast to the
real part of binding energy, the imaginary part of the binding energy 
increases with the temperature but increases with the anisotropy. With 
these observations, we have now computed the dissociation temperatures 
at different anisotropies (Table 1), where $J/\psi$ is dissociated at 
$2.46~T_c$ and $2.49~T_c$ for $\xi=0.3$ and $0.6$, respectively 
(obtained from the intersection of binding energies) whereas 
$\Upsilon$'s are dissociated at $3.45~T_c$ and $3.46T_c$, respectively.
Thus the presence of anisotropy 
enhances the dissociation point to the resonances.
Like in isotropic medium, we also computed the dissociation temperatures
from the criterion on the thermal width and found the temperatures become smaller.
For example, $J/\psi$ is now dissociated at $1.46~T_c$ and $1.54~T_c$ 
and $\Upsilon$ is dissociated at $3.17T_c$ and $3.26~T_c$, for
the same anisotropies.
% so they seem to compensate to each other and the
%dissociation temperature remains unchanged. 
%But since the contribution
%due to the imaginary part is much smaller than the
%contribution due to the real part so the effect of anisotropy on the
%real part wins over the imaginary part, thus results in increase of
%dissociation temperature with the anisotropy.

\section{Conclusion}
We have investigated the properties of charmonium and bottomonium states
through the in-medium modifications to both perturbative and nonperturbative
of the Cornell potential, not its perturbative term alone as usually done
in the literature. For this purpose we have obtained both the real and
imaginary part of the potential within the framework of real-time
formalism, in both isotropic and anisotropic medium. In isotropic medium,
the inclusion of the linear/string term, in addition to the Coulomb term,
makes the real part of the potential more attractive. So, as a consequence
the quarkonium states become more bound compared to the medium modification
to the Coulomb term alone. Moreover the string term affects the
imaginary part too where its magnitude is increased by the string
contribution. As a result, the (thermal) width of the states are
broadened due to the presence of string term and makes the competition
between the screening and the broadening due to damping interesting
and plays an important role in the dissociation mechanism.
With these cumulative observations, we studied the dissociation in a 
medium where a resonance is said to be dissolved in a 
medium~\cite{Laine:2007qy,Burnier:2009yu} either when its (real) binding
energy decreases with temperature and becomes equal to its width or
the real and imaginary binding energy becomes equal.
We have found that the quarkonium states are dissociated at higher temperature
compared to the medium-consideration of the Coulomb term only.

We have then extended our exploration of quarkonium to a medium which 
exhibits a local
anisotropy in the momentum space. This may arise due to the rapid
expansion in the beam direction compared to its transverse direction,
at the early stage of the evolution in ultra-relativistic heavy-ion
collisions. For that, we have first revisited the anisotropic corrections to the
retarded, advanced and symmetric propagators through their self-energies
in the hard-loop resummation technique and apply these results to
calculate the medium-corrections to the perturbative and nonperturbative
term of the Cornell potential. We are however restricted to a medium closer
to equilibrium/isotropic because although the system was initially 
anisotropic but by 
the time quarkonium resonances are formed in plasma ($t_F=\gamma \tau_F$, $\tau_F$
is the formation time in the rest frame of quarkonium), the plasma becomes
almost isotropic.

The effect of nonvanishing nonperturbative term on the quarkonium properties,
as seen earlier, remains the same even in the presence of momentum
anisotropy. However, the anisotropy behaves as an additional handle to
decipher the properties of quarkonium states, namely, in anisotropic
medium, the binding of $Q \bar Q$ pairs gets
stronger with respect to their isotropic counterpart because
both the real and imaginary part of the complex potential becomes 
deeper with the increase of anisotropy. This
is due to the fact that the (effective) Debye mass in anisotropic
medium is always smaller than in isotropic medium. As a result
the screening of the Coulomb and string contributions is
less accentuated and thus quarkonium states
are bound more strongly than in isotropic medium.
%On the other hand, the effect of anisotropy on the imaginary part
%of the potential is opposite to the real part, hence the width of the
%resonances become smaller compared to their isotropic counterpart, with
%the increase of anisotropy. 
The overall observation is that the dissociation
temperature increases with the increase of anisotropy. For example,
$ J/\psi $ is dissociated at $2.45~T_c$, $2.46~T_c$, and $2.49~T_c$ for 
the anisotropies $\xi=$ 0, 0.3, and 0.6, respectively.
Similarly, $\Upsilon$ is dissociated at $3.40~T_c$, $3.45~T_c$, and 
$3.46~T_c$ for $\xi=0$, 0.3, and 0.6, respectively.

Our results on the dissociation temperatures are found relatively higher compared to similar
calculation~\cite{Strickland:2011aa,Margotta:2011ta}, which may be due to the absence of
three-dimensional medium modification of the linear term in their
calculation. In fact, one-dimensional Fourier transform of the Cornell
potential yields the similar form used in the lattice QCD in which
one-dimensional color flux tube structure was assumed~\cite{dixit}.
However, at finite temperature that may not be the case since the
flux-tube structure may expand in more dimensions~\cite{Satz}.
Therefore, it would be better to consider the three-dimensional form of
the medium modified Cornell potential which has been done exactly in
the present work.

In brief, the properties of quarkonium states are affected 
by the inclusion of the non-perturbative (string) term in the 
potential, in addition to the anisotropic medium effects, which
needs more careful treatment due to its nonperturbative character 
in future. 

\noindent {\bf Acknowledgments:}
We are thankful for some financial assistance from CSIR project (CSR-656-PHY),
Government of India. 

\end{document}